\documentclass{jfm}

\usepackage{graphicx}
\usepackage{newtxtext}
\usepackage{newtxmath}
\usepackage{natbib}
\usepackage{hyperref}
\usepackage{amsmath}
\DeclareMathOperator{\erf}{erf}
\DeclareMathOperator{\erfc}{erfc}

\usepackage[normalem]{ulem}

\usepackage{nth}
\hypersetup{
    colorlinks = true,
    urlcolor   = blue,
    citecolor  = black,
}

\newcommand{\RomanNumeralCaps}[1]
\linenumbers

\newif\ifrevmode
\revmodetrue          

\title{Cross-streamline diffusiophoretic migration of colloids in Taylor-dispersed channel flows}

\author{Yiran Li\aff{1},
  Mobin Alipour\aff{1}
 \and Amir A. Pahlavan\aff{1}\corresp{\email{amir.pahlavan@yale.edu}}}

\affiliation{\aff{1} Mechanical Engineering and Materials Science,\\
 Yale University, New Haven, Connecticut 06511, USA.}

\begin{document}
\maketitle

\begin{abstract}
Diffusiophoretic transport of colloids in pressure-driven channel flow is
commonly analysed in two limits: an early-time regime in which the solute field
is fully two-dimensional, and a late-time macrotransport regime in which
cross-sectional homogenization leaves only a weak axial bias on the particles.
For colloids, however, many experiments operate in the broad intermediate
window \(a^2/D_{\mathrm s}\ll t\ll a^2/D_{\mathrm p}\): the solute has entered
the Taylor-dispersion regime, but the particles remain effectively
non-diffusive across the gap. We show that the Taylor-dispersed solute retains
a residual transverse gradient that is P\'eclet-enhanced relative to the axial
gradient and decays only as \(t^{-1/2}\). This gradient is small in the solute
concentration but large enough in \(\nabla\ln c\) to drive cross-streamline
migration of colloids. Attractive fronts (\(c_{\mathrm f}>c_{\mathrm i}\))
move particles toward faster centreline streamlines, sharpening the leading
edge and accelerating removal; repulsive fronts (\(c_{\mathrm f}<c_{\mathrm i}\))
move particles toward slower near-wall streamlines, broadening the trailing
edge and delaying removal. Direct simulations and microfluidic experiments
confirm these front-sharpening and front-broadening dynamics. An asymptotic
Taylor-regime solute field, combined with a non-diffusive trajectory model,
captures the observed front geometries, density profiles, and removal
dynamics. The results show that Taylor-dispersed solute fields can remain
dynamically two-dimensional for particles, even when their concentration is
nearly cross-sectionally uniform.
\end{abstract}

\begin{keywords}
Authors should not enter keywords on the manuscript, as these must be chosen by the author during the online submission process and will then be added during the typesetting process (see \href{https://www.cambridge.org/core/journals/journal-of-fluid-mechanics/information/list-of-keywords}{Keyword PDF} for the full list).  Other classifications will be added at the same time.
\end{keywords}

{\bf MSC Codes }  {\it(Optional)} Please enter your MSC Codes here

\section{\label{sec:intro}Introduction}

Solute concentration gradients can drive the directed migration of colloidal
particles by diffusiophoresis \citep{anderson1989colloid,prieve1984motion,velegol2016origins,Marbach19,shim2022diffusiophoresis,ault2025physicochemical}. This coupling between solute transport and particle
motion underlies a wide range of natural and engineered processes, including
particle focusing and separation in microfluidic devices, membrane fouling and
cleaning, transport in biological materials and biofilms, contaminant transport
in groundwater, and enhanced recovery from porous formations
\citep{shi2016diffusiophoretic,shin2016size,shin2017accumulation,shin2018cleaning,shim2022diffusiophoresis,ault2025physicochemical,Pahlavan26}.
At first sight, however, diffusiophoresis would appear to have only a weak
influence on particles carried by pressure-driven channel flows. Molecular
solutes with diffusivity \(D_{\mathrm s}\) diffuse rapidly across a channel of half-width \(a\), over a time
\(a^2/D_{\mathrm s}\), so the solute field is often treated as effectively
one-dimensional once this transverse equilibration has occurred, with gradients
only along the flow direction \citep{chu2021macrotransport}. Since phoretic velocities are also much
smaller than the typical background advective velocities, this reasoning suggests that
diffusiophoresis should produce only a weak axial bias in flowing channels.

Recent observations on diffusiophoretic colloid transport in porous media
challenge this expectation \citep{jotkar2024impact,alipour2026diffusiophoretic,Pujari26}. Solute
gradients substantially altered colloid travel times and breakthrough behaviour
within preferential flow pathways, where flow velocities exceed phoretic
velocities by orders of magnitude. Strikingly, similar signatures appeared
whether dead-end pores were initially filled with particles or left empty,
suggesting that phoretic exchange with stagnant pockets was not the dominant
mechanism \citep{alipour2026diffusiophoretic}. The observations instead pointed to cross-streamline migration of
colloids within the flowing pathways themselves. This raises a basic question:
how can solute gradients drive appreciable transverse particle motion in a
channel after the solute has apparently reached the Taylor-dispersion regime?

Diffusiophoretic transport in channel flows has previously been analysed in
two limiting regimes. At early times, before solute diffusion has equilibrated
across the cross-section, the solute and particle fields are fully
two-dimensional \citep{migacz2022diffusiophoresis}. At late times, motivated
by classical Taylor--Aris dispersion theory
\citep{taylor1953dispersion,aris1956dispersion,Stone04,Squires05}, both fields
can be cross-sectionally averaged, leading to a one-dimensional macrotransport
description in which axial solute gradients generate an effective advective
bias for the particles \citep{chu2021macrotransport}. Related Taylor-style
reductions have also been developed for diffusioosmotic flows \citep{alessio2022diffusioosmosis,teng2023diffusioosmotic}. Within this framework, axial gradients
alone cannot produce cross-streamline redistribution, and
diffusiophoresis appears only as a weak correction to the mean particle
velocity. This conclusion would imply that
diffusiophoresis is essentially irrelevant in flowing channels at the velocities
and times relevant to typical experiments, in conflict with the porous-media
observations above.

These two limits are widely separated for colloids because the particle
diffusivity \(D_{\mathrm p}\) is typically much smaller than the solute
diffusivity \(D_{\mathrm s}\). Denoting by \(a\) a characteristic transverse
length scale of the channel, this separation gives rise to a broad
intermediate regime,
\[
a^2/D_{\mathrm s}
\ll t \ll
a^2/D_{\mathrm p},
\]
in which the solute has entered the Taylor-dispersion regime, whereas the
particle distribution has not yet homogenized across the cross-section. This
intermediate regime is the focus of the present work.

The key point of this paper is that the Taylor-dispersed solute field is not
exactly one-dimensional at the level relevant for diffusiophoresis. Although
the leading-order solute concentration is uniform across
the channel, shear maintains a transverse correction that scales as
\[
\partial_y c \sim \mathrm{Pe}_{\mathrm s}\,\partial_x c,
\qquad
\mathrm{Pe}_{\mathrm s}=\frac{U_{\mathrm m}a}{D_{\mathrm s}},
\]
where \(c(x,y,t)\) is the solute concentration, \(U_{\mathrm m}\) is the cross-sectionally averaged flow velocity
and \(\mathrm{Pe}_{\mathrm s}\) is the solute P\'eclet number. This correction
decays only as \(t^{-1/2}\) and, although small in the solute concentration, produces a transverse gradient comparable to or larger than the axial gradient. Because
diffusiophoresis depends on \(\nabla\ln c\), and because the colloids are
effectively non-diffusive over the times of interest, this residual transverse
gradient can drive appreciable cross-streamline migration. Thus a correction
that is weak in \(c\) but not in \(\nabla c\) can redistribute particles across
the Poiseuille profile and change their travel times and removal dynamics.

Here we combine two-dimensional simulations, microfluidic experiments,
asymptotic theory, and a reduced trajectory-based model to show that this
intermediate regime supports a distinct mode of diffusiophoretic transport.
Attractive and repulsive solute fronts produce different particle-front shapes
and markedly different removal dynamics, despite the solute field being nearly
cross-sectionally uniform throughout. We derive the Taylor-regime solute field
including its transverse correction and use it to construct a non-diffusive
model that captures the front shapes, density profiles, and removal dynamics
observed in simulations and experiments. The results identify a transport
regime in which a nominally small correction to the solute field has a
leading-order effect on colloid transport, and they help reconcile the apparent
discrepancy between channel-flow theory and observations in porous media,
filtration, and other confined transport settings.

\section{\label{sec:Simulation}Diffusiophoretic particle transport in a two-dimensional channel}

To demonstrate that residual transverse solute gradients can strongly modify
particle transport, we perform two-dimensional numerical simulations using
complementary Lagrangian and Eulerian descriptions of the particle phase. In
both formulations, the solute concentration is obtained by solving the full
advection--diffusion equation. In the Lagrangian simulations, individual
non-diffusive particles are advected by the combined flow and
diffusiophoretic velocities. In the Eulerian simulations, the particle phase
is represented by a density field governed by an advection--diffusion equation
with finite particle diffusivity. Further details of the numerical discretization and particle-tracking procedure are provided in  Appdendix~\ref{app:numerical_methods}.

Both approaches retain the full two-dimensional solute field and therefore do
not invoke the Taylor-regime approximation developed in
\S\ref{sec:Model}. The Eulerian simulations additionally resolve finite
particle diffusion, whereas the Lagrangian simulations isolate the
non-diffusive transport mechanism underlying the reduced model. Together, they
provide complementary benchmarks for the theoretical description developed
below.

\subsection{\label{sec:Simulation_setup}Problem setup}

We consider solute and colloid transport in a two-dimensional channel of
length \(L\) and full width \(2a\). The streamwise coordinate is \(x\), and
the transverse coordinate \(y\) is measured from the centreline. Because the
geometry, flow, and imposed conditions are symmetric about \(y=0\), only the
upper half-channel,
\[
0\leq x\leq L,
\qquad
0\leq y\leq a,
\]
is resolved numerically.

The imposed flow is the steady plane-Poiseuille profile
\begin{equation}
\mathbf{u}(y)
=
\frac{3}{2}U_{\mathrm m}
\left(
1-\frac{y^2}{a^2}
\right)\hat{\mathbf{x}},
\label{eq:poiseuille_sim}
\end{equation}
where \(U_{\mathrm m}\) is the cross-sectionally averaged velocity. The solute
concentration \(c(x,y,t)\) satisfies
\begin{equation}
\partial_t c+\mathbf{u}\cdot\nabla c
=
D_{\mathrm s}\nabla^2 c,
\label{eq:c_ade}
\end{equation}
where \(D_{\mathrm s}\) is the solute diffusivity. Symmetry at the centreline
and impermeability of the wall give
\begin{equation}
\partial_y c=0
\qquad
\text{at}
\qquad
y=0,a.
\label{eq:c_transverse_bc}
\end{equation}

We use complementary Lagrangian and Eulerian descriptions of the particle
phase. In the Lagrangian simulations, the position
\(\mathbf{R}_k(t)=(X_k(t),Y_k(t))\) of particle \(k\) evolves according to
\begin{equation}
\frac{\mathrm d\mathbf{R}_k}{\mathrm dt}
=
\mathbf{u}\bigl(Y_k(t)\bigr)
+
\mathbf{u}_{\mathrm{DP}}\bigl(\mathbf{R}_k(t),t\bigr),
\label{eq:lagrangian_trajectory}
\end{equation}
where
\begin{equation}
\mathbf{u}_{\mathrm{DP}}
=
\Gamma_{\mathrm p}\nabla\ln c
\label{eq:uDP}
\end{equation}
is the diffusiophoretic velocity and \(\Gamma_{\mathrm p}\) is the particle
mobility. No stochastic Brownian-displacement term is included in
\eqref{eq:lagrangian_trajectory}; the Lagrangian particles are therefore
strictly non-diffusive.

In the Eulerian simulations, the particle phase is represented by a number
density \(n(x,y,t)\), which satisfies
\begin{equation}
\partial_t n
+
\nabla\cdot
\left[
\left(
\mathbf{u}+\mathbf{u}_{\mathrm{DP}}
\right)n
\right]
=
D_{\mathrm p}\nabla^2 n,
\label{eq:n_ade}
\end{equation}
where \(D_{\mathrm p}\) is the particle diffusivity. The transverse boundary conditions are
\begin{equation}
\left[\left(
\mathbf{u}+\mathbf{u}_{\mathrm{DP}}
\right)n
-
D_{\mathrm p}\nabla n\right]\cdot\hat{\boldsymbol{\nu}}=0
\qquad
\text{at}
\qquad
y=0,a,
\label{eq:n_transverse_bc}
\end{equation}
where \(\hat{\boldsymbol{\nu}}\) is the normal to the boundary of the
computational half-channel. Equation~\eqref{eq:n_transverse_bc} represents
symmetry at the centreline and zero normal particle flux at the wall.

In both simulation protocols, the channel initially contains solute at
concentration \(c_{\mathrm i}\). At \(t=0\), the inlet concentration is
switched to \(c_{\mathrm f}\), so that
\begin{equation}
c(x,y,0)=c_{\mathrm i},
\qquad
c(0,y,t)=c_{\mathrm f}
\quad
\text{for}
\quad
t>0.
\label{eq:c_initial_inlet}
\end{equation}
A convective outflow condition is imposed at \(x=L\). In the Eulerian
simulations, the channel is initially filled with a uniform colloidal
suspension,
\begin{equation}
n(x,y,0)=n_0,
\label{eq:n_initial}
\end{equation}
whereas the incoming solution is particle-free,
\begin{equation}
n(0,y,t)=0
\qquad
\text{for}
\qquad
t>0.
\label{eq:n_inlet}
\end{equation}
Thus, the solute and particle fronts are generated simultaneously at the
inlet.

We consider three solute-displacement protocols: a control case,
\(c_{\mathrm f}/c_{\mathrm i}=1\); an attractive case,
\(c_{\mathrm f}/c_{\mathrm i}=100\); and a repulsive case,
\(c_{\mathrm f}/c_{\mathrm i}=0.01\). The terms ``attractive'' and
``repulsive'' describe the direction of particle migration relative to the
invading solute front for positive mobility,
\(\Gamma_{\mathrm p}>0\).

Unless otherwise stated, the transport parameters are
\(D_{\mathrm s}=10^{-9}\,\mathrm{m^2\,s^{-1}}\), \(D_{\mathrm p}=10^{-13}\,\mathrm{m^2\,s^{-1}}\), and \(
\Gamma_{\mathrm p}=0.6\times10^{-9}\,\mathrm{m^2\,s^{-1}}.\)
The channel dimensions and mean velocity are
\(a=20\,\mu\mathrm{m}\), \(L=30\,\mathrm{mm}\), and \(U_{\mathrm m}=100\,\mu\mathrm{m\,s^{-1}}\).
The solute and particle P\'eclet numbers based on the channel half-width are
therefore
\begin{equation}
\mathrm{Pe}_{\mathrm s}
=
\frac{U_{\mathrm m}a}{D_{\mathrm s}}
=
2,
\qquad
\mathrm{Pe}_{\mathrm p}
=
\frac{U_{\mathrm m}a}{D_{\mathrm p}}
=
2\times10^4.
\label{eq:simulation_pe}
\end{equation}
The corresponding transverse diffusion and advective residence times are
\begin{equation}
\tau_{\mathrm s}
=
\frac{a^2}{D_{\mathrm s}}
=
0.4\,\mathrm{s}
\;\ll\;
\tau_{\mathrm{PV}}
=
\frac{L}{U_{\mathrm m}}
=
300\,\mathrm{s}
\;\ll\;
\tau_{\mathrm p}
=
\frac{a^2}{D_{\mathrm p}}
=
4\times10^3\,\mathrm{s}.
\label{eq:timescales}
\end{equation}
Here \(\tau_{\mathrm{PV}}\) is the advective residence, or pore-volume, time
of the channel. The Eulerian simulations therefore lie in the intermediate
regime introduced in \S\ref{sec:intro}: the solute equilibrates across the
channel much faster than it traverses the channel, whereas particle diffusion
is too slow to homogenize the particle distribution across the gap over one
residence time. Particle diffusion is nevertheless retained in
\eqref{eq:n_ade} so that its local smoothing effect can be assessed.

The Eulerian and Lagrangian calculations use distinct particle
initializations because they serve complementary purposes. The Eulerian
simulations begin with the simultaneous solute and particle displacement
specified by \eqref{eq:c_initial_inlet}--\eqref{eq:n_inlet}. This
initialization preserves the coincident solute and particle fronts used in the
reduced initial-value problem of \S\ref{sec:Model}, thereby facilitating a
direct comparison between the full numerical solution and the reduced model.

The Lagrangian simulations are instead designed to isolate particle transport
after the early transverse-equilibration stage of the solute has passed. We
first evolve the solute field in the absence of particles for $t_{\mathrm{pre}}=20\,\mathrm{s}=50\tau_{\mathrm s}$. At \(t=t_{\mathrm{pre}}\), particles are seeded uniformly in a region ahead
of, and sufficiently far downstream from, the solute transition that the diffusiophoretic drift is negligible at the instant of release. The
particles are subsequently evolved using
\eqref{eq:lagrangian_trajectory}, with particle diffusion neglected. The
geometry, flow, solute-transport parameters, and concentration ratios are
otherwise identical to those used in the Eulerian simulations.

\begin{figure*}
    \centering
    \includegraphics[
        width=\textwidth,
        keepaspectratio
    ]{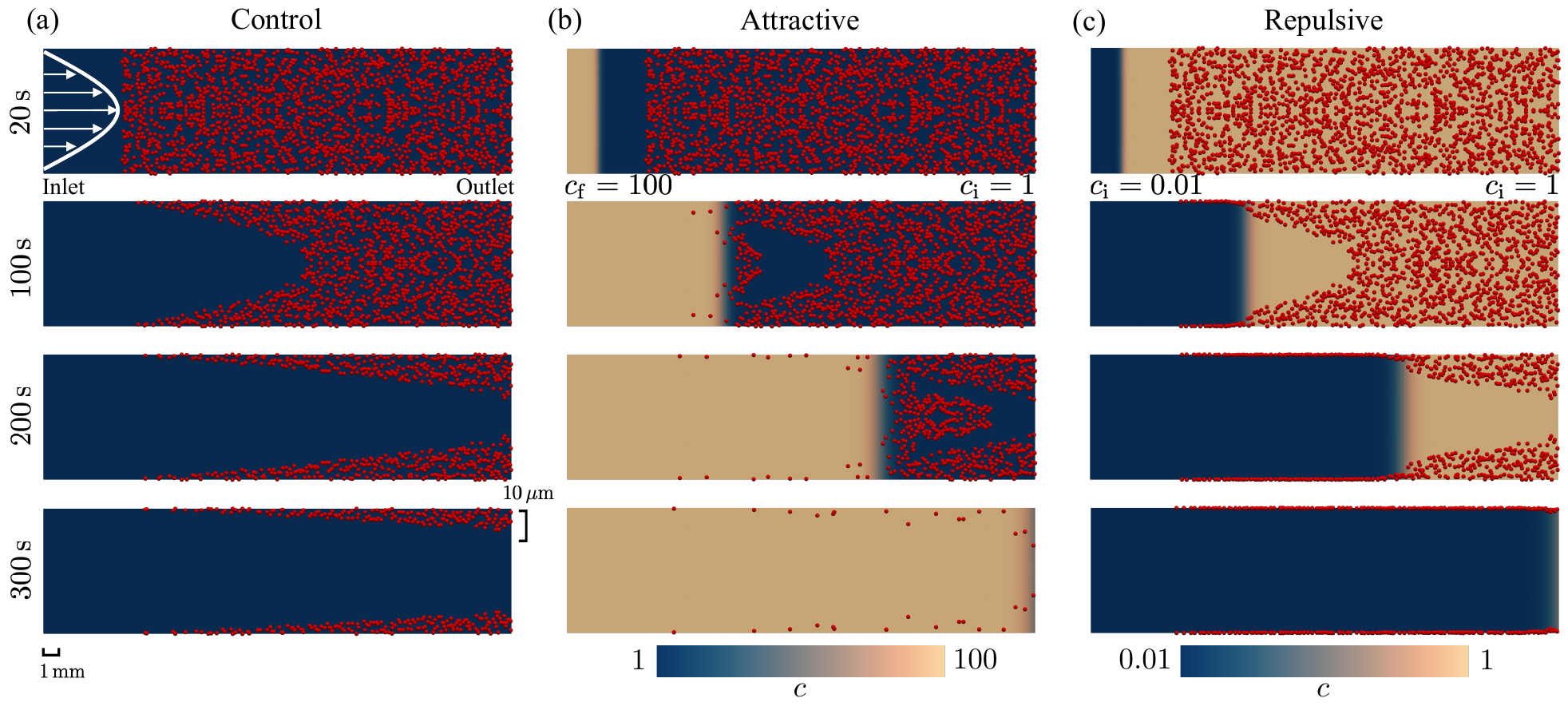}
    \captionsetup{
        width=\textwidth,
        justification=justified,
        singlelinecheck=false
    }
    \caption{
    Lagrangian particle-tracking simulations for
    (a) the control case, \(c_{\mathrm f}/c_{\mathrm i}=1\);
    (b) the attractive case, \(c_{\mathrm f}/c_{\mathrm i}=100\); and
    (c) the repulsive case, \(c_{\mathrm f}/c_{\mathrm i}=0.01\).
    Red markers denote the instantaneous particle positions, and the
    background shading denotes the solute concentration. The rows correspond
    to \(t=\{20,\,100,\,200,\,300\}\,\mathrm{s}\), measured from the onset of
    the solute displacement. The solute field is first evolved without
    particles until \(t_{\mathrm{pre}}=20\,\mathrm{s}\), at which time the
    particles are released uniformly in a downstream region where the initial
    solute gradients are negligible. The first row therefore shows the
    particle configuration at release, while the subsequent rows correspond
    to elapsed particle-tracking times of \(80\), \(180\), and
    \(280\,\mathrm{s}\), respectively. Only the upper half-channel is resolved
    computationally; for visualization, the solute and particle fields are
    reflected about the centreline. The streamwise and transverse coordinates
    are displayed at different magnifications, as indicated by the separate
    \(1\,\mathrm{mm}\) and \(10\,\mu\mathrm{m}\) scale bars in panel (a).
    The imposed Poiseuille velocity profile and flow direction are indicated
    schematically in the upper-left panel.
    }
    \label{fig:LPT}
\end{figure*}

\subsection{\label{sec:Simulation_results}Diffusiophoresis reshapes particle fronts}

\begin{figure*}
    \centering
    \includegraphics[
        width=\textwidth,
        keepaspectratio
    ]{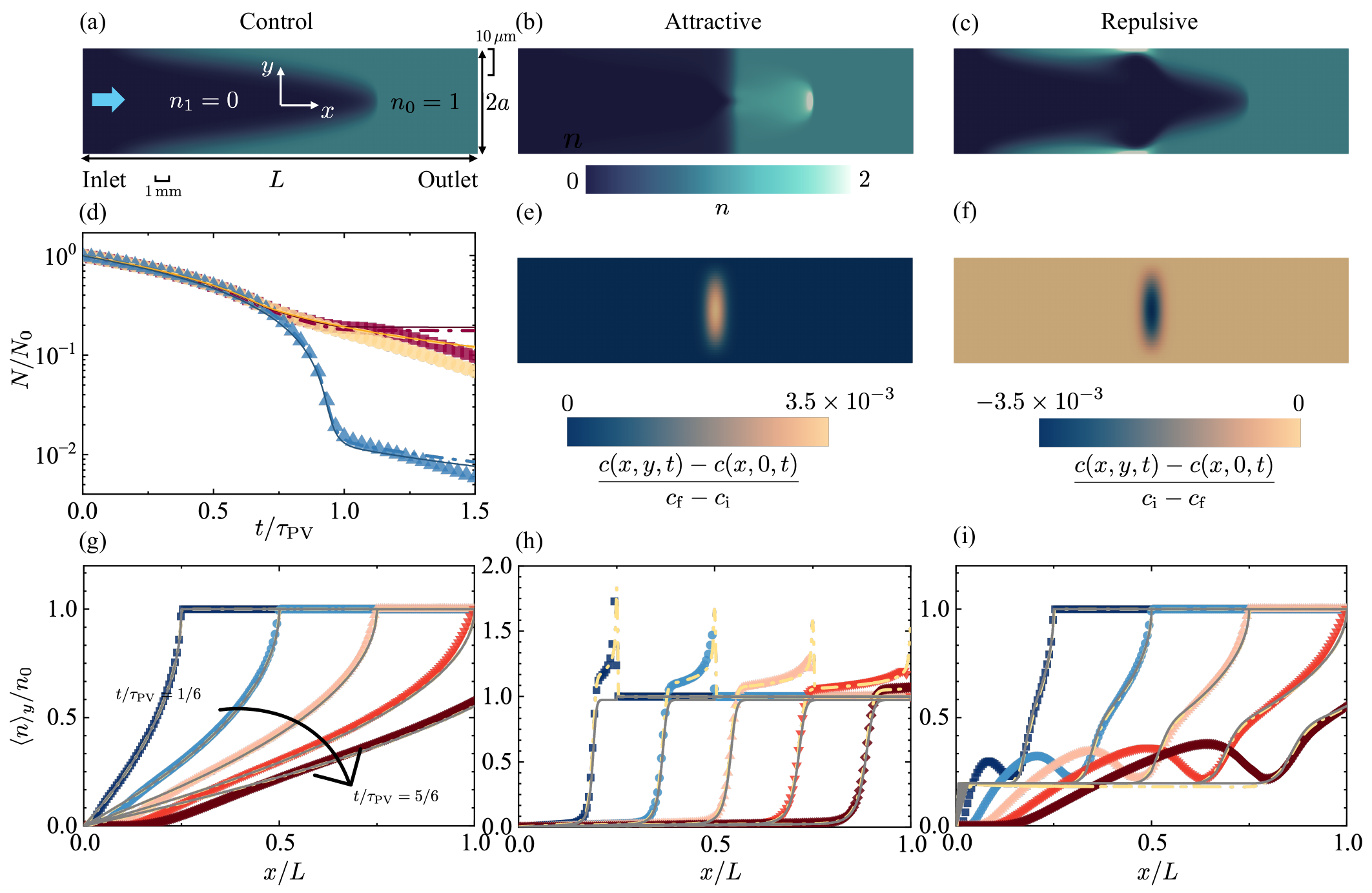}
    \captionsetup{
        width=\textwidth,
        justification=justified,
        singlelinecheck=false
    }
    \caption{
    Influence of diffusiophoresis on particle transport in a
    two-dimensional channel. Only the upper half-channel is resolved
    numerically; in the field panels, the numerical solutions are reflected
    about the centreline to display the full channel. The streamwise and
    transverse directions in panels (a--c,e,f) are shown at different
    magnifications for clarity.
    The channel length and full width are denoted by \(L\) and \(2a\),
    respectively, and the arrow in panel (a) indicates the flow direction.
    In panels (d,g--i), symbols denote the full particle
    advection--diffusion simulations of \eqref{eq:n_ade}, dash-dotted curves
    denote the non-diffusive advection model introduced in
    \S\ref{sec:Model}, and solid curves denote the analytical predictions
    developed in \S\ref{sec:Model}.
    (a--c) Particle density \(n(x,y,t)\), normalized by setting the initially
    uniform density to \(n_0=1\), at
    \(t/\tau_{\mathrm{PV}}=1/2\) for
    (a) the control case, \(c_{\mathrm f}/c_{\mathrm i}=1\);
    (b) the attractive case, \(c_{\mathrm f}/c_{\mathrm i}=100\); and
    (c) the repulsive case, \(c_{\mathrm f}/c_{\mathrm i}=0.01\).
    (d) Fraction \(N(t)/N_0\) of the initial particle population remaining
    in the channel for the control (yellow), attractive (blue), and repulsive
    (red) cases.
    (e,f) Centreline-referenced transverse solute-concentration deviations
    at \(t/\tau_{\mathrm{PV}}=1/2\), shown as
    \(\bigl[c(x,y,t)-c(x,0,t)\bigr]/(c_{\mathrm f}-c_{\mathrm i})\)
    for the attractive case (e) and
    \(\bigl[c(x,y,t)-c(x,0,t)\bigr]/(c_{\mathrm i}-c_{\mathrm f})\)
    for the repulsive case (f). The control case has
    \(c_{\mathrm f}=c_{\mathrm i}\), so the transverse concentration
    deviation vanishes identically and is omitted.
    (g--i) Normalized cross-sectionally averaged particle density
    \(\langle n\rangle_y/n_0\) at
    \(t/\tau_{\mathrm{PV}}
    =\{1/6,\,1/3,\,1/2,\,2/3,\,5/6\}\)
    for the control (g), attractive (h), and repulsive (i) cases. Colours
    progress from blue to red with increasing time. The \(x/L\) axes in
    panels (g--i) are horizontally aligned with the corresponding field
    panels (a--c), allowing the full-simulation profiles at
    \(t/\tau_{\mathrm{PV}}=1/2\) to be compared directly with the
    two-dimensional particle fields at the same streamwise locations.
    }
    \label{fig:Simulation}
\end{figure*}

Figure~\ref{fig:LPT} provides a direct test of whether diffusiophoretic
reshaping persists after the initial transverse equilibration of the solute.
The solute field is first evolved without particles until
\(t_{\mathrm{pre}}=20\,\mathrm{s}=50\tau_{\mathrm s}\), after which
non-diffusive particles are released in a downstream region ahead of the
solute front, where solute gradients are initially negligible.
Consequently, any subsequent cross-streamline migration is produced when the
already Taylor-dispersed solute front reaches the particles and cannot be
attributed to the initial \(t=O(\tau_{\mathrm s})\) equilibration transient.

In the control case, \(c_{\mathrm f}/c_{\mathrm i}=1\), so that
\(\mathbf{u}_{\mathrm{DP}}=\mathbf{0}\). The particle interface is then
deformed solely by Poiseuille flow: particles near the centreline are
transported more rapidly than those near the wall, leaving an extended
near-wall tail. In the attractive case,
\(c_{\mathrm f}/c_{\mathrm i}=100\), the residual transverse solute gradient
drives particles toward the centreline. Particles initially located on slower
off-centre streamlines are therefore transferred to faster central
streamlines. This migration suppresses the population that would otherwise
lag behind the solute front and produces a compact particle distribution. In the repulsive case,
\(c_{\mathrm f}/c_{\mathrm i}=0.01\), the transverse drift is reversed.
Particles migrate toward the wall, where they form narrow particle-rich bands
and are transported by the slowest part of the Poiseuille profile. The
resulting near-wall retention generates a broad trailing distribution. The
opposite deformations in the attractive and repulsive cases remain evident
long after particle release, demonstrating that residual Taylor-regime
gradients can reorganize particle transport over the full channel width.

The same migration signatures are recovered in the Eulerian simulations. At
\(t/\tau_{\mathrm{PV}}=1/2\), the control field in
figure~\ref{fig:Simulation}(a) displays the familiar shear-stretched particle
interface generated by the Poiseuille velocity profile. In the attractive
case, figure~\ref{fig:Simulation}(b), migration toward the centreline produces
a much more compact streamwise transition together with a localized
particle-density enhancement near the centreline. In the repulsive case,
figure~\ref{fig:Simulation}(c), particles are displaced toward the wall,
forming particle-rich near-wall layers that extend well behind the central
portion of the front.

The cross-sectionally averaged profiles in
figure~\ref{fig:Simulation}(g--i) quantify these differences. Because their
\(x/L\) axes are horizontally aligned with the fields in
figure~\ref{fig:Simulation}(a--c), the full-simulation profiles at
\(t/\tau_{\mathrm{PV}}=1/2\) can be compared directly with the corresponding
two-dimensional particle distributions. The control profile broadens smoothly
under Poiseuille shear. The attractive profile instead contains a sharp
transition and a localized density overshoot above \(n_0\), corresponding to
the concentrated region visible in figure~\ref{fig:Simulation}(b). The
repulsive profile develops an extended, non-monotonic low-density tail,
reflecting the population retained in the slow-moving near-wall layers.
The profiles at the other times show that these distinct front geometries
persist throughout the displacement.

Panels~\ref{fig:Simulation}(e,f) expose the weak solute structure responsible
for this large particle response. These panels display the centreline-referenced transverse
deviation \(c(x,y,t)-c(x,0,t)\). The deviations are localized around the
Taylor-dispersed solute front, have opposite signs in the attractive and
repulsive cases, and have normalized magnitudes of only
\(O(10^{-3})\).
Thus, on the scale of the imposed streamwise concentration difference, the
solute remains nearly uniform across the channel, but its transverse
correction is not zero. Because diffusiophoresis depends on
\(\nabla\ln c\), this small concentration difference across the narrow
channel gap generates a coherent transverse drift that transfers particles
between streamlines and produces an \(O(1)\) change in their streamwise
distribution. A solute field that is nearly one-dimensional in concentration
can therefore remain dynamically two-dimensional for the particles.

This streamline redistribution directly changes particle removal from the
channel. As shown in figure~\ref{fig:Simulation}(d), the residual particle
fraction \(N(t)/N_0\) decreases most rapidly in the attractive case. Migration
toward the centreline transfers particles onto faster streamlines, causing the
compact particle front to leave the channel rapidly as it approaches one
residence time. The repulsive case exhibits the slowest removal and develops a
pronounced shoulder or plateau after approximately one residence time,
consistent with prolonged retention in the slow-moving near-wall region. The
control response lies between these two limits. The microscopic
cross-streamline migration therefore produces a substantial change in the
channel-scale particle residence and clearance dynamics.

The dash-dotted non-diffusive calculations and solid analytical predictions
included in figure~\ref{fig:Simulation}(d,g--i) are developed in
\S\ref{sec:Model} and assessed quantitatively in
\S\ref{subsec:Validation}. The relevant observation here is that removing
particle diffusion does not eliminate the compact attractive front, the broad
repulsive tail, or their ordering in the particle-removal curves. Finite
particle diffusion primarily rounds sharp interfaces, reduces localized
density peaks, and smooths the near-wall layers generated by the
non-diffusive dynamics. Together with the strictly non-diffusive Lagrangian
simulations, performed after the initial solute-equilibration stage, these
results show that neither particle diffusion nor the initial
\(t=O(\tau_{\mathrm s})\) solute transient is required to produce the observed
redistribution. The dominant mechanism is cross-streamline diffusiophoretic
advection generated by the residual transverse gradient of the
Taylor-dispersed solute front. This observation motivates the reduced
trajectory-based description developed in \S\ref{sec:Model}.

\section{\label{sec:Experiments}Microfluidic experiments}

The simulations of \S\ref{sec:Simulation} predict that attractive and
repulsive solute fronts produce distinct particle-front shapes during
displacement in channel flow. We now show that the same signatures are
observed in microfluidic experiments. The purpose of this section is to test
whether the cross-streamline migration mechanism identified numerically is
observable in a real channel-flow geometry. The analytical reduction that
explains these signatures is developed in \S\ref{sec:Model}.

\begin{figure*}
    \centering
    \includegraphics[width=\textwidth,keepaspectratio]{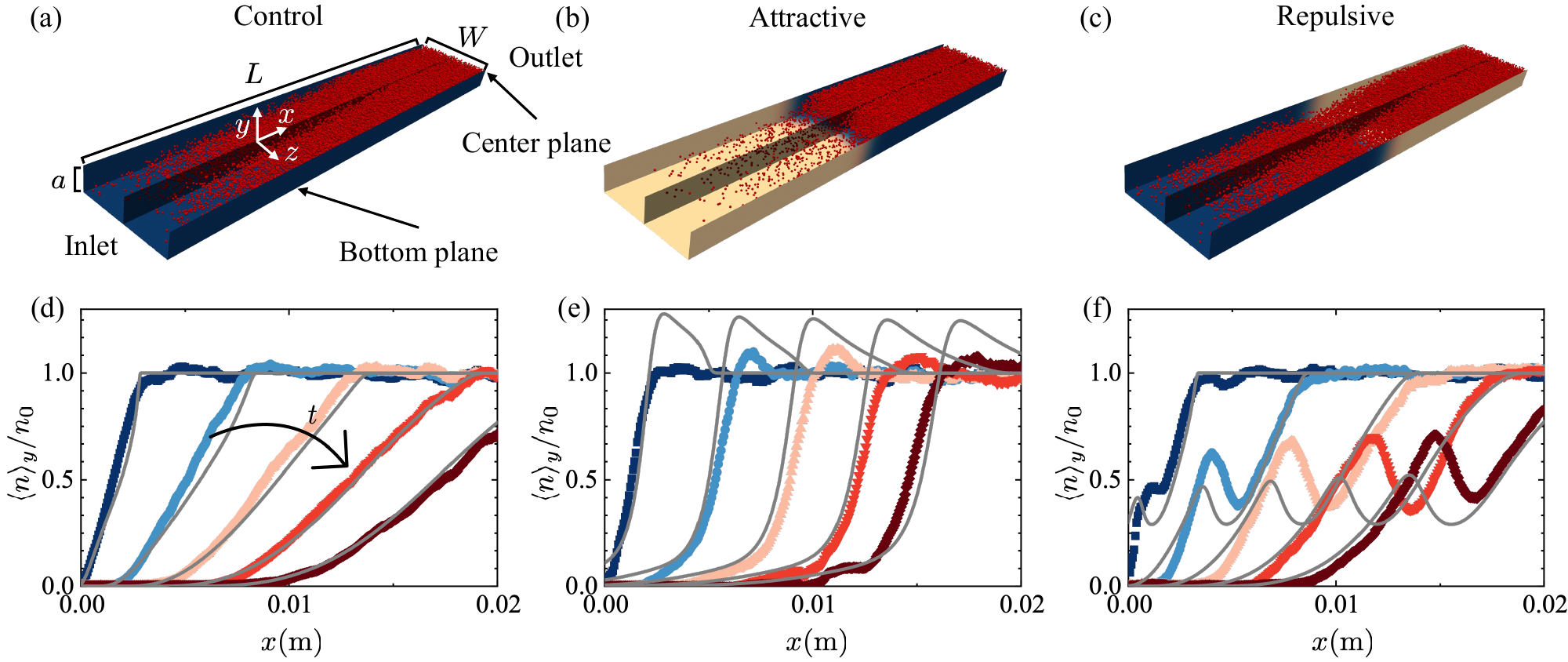}
    \captionsetup{
        width=\textwidth,
        justification=justified,
        singlelinecheck=false
    }
    \caption{
    Microfluidic displacement experiments in a straight rectangular channel
    with gap thickness \(2a=25\,\mu\mathrm{m}\), width
    \(W=5\,\mathrm{mm}\), and length \(L=25\,\mathrm{mm}\).
    (a--c) Schematics, not to scale, of the experimental configurations for
    (a) the control case, \(\beta=1\);
    (b) the attractive case, \(\beta=100\); and
    (c) the repulsive case, \(\beta=0.01\), where
    \(\beta=c_{\mathrm f}/c_{\mathrm i}\).
    Only the lower half of the channel gap is shown. Red points indicate the colloidal particles.
    The inset \(x\)--\(y\) section at the spanwise midplane \(z=0\)
    corresponds to the two-dimensional channel considered in the simulations. In panels (b,c), the wall shading denotes the instantaneous
    solute concentration and indicates the position of the solute front.
    Because \(W\gg 2a\), spanwise variations are negligible near \(z=0\).
    Top-view fluorescence images project the particle detections through the
    gap direction \(y\); averaging the detections over a central spanwise region
    gives the streamwise particle-density profiles shown in panels (d--f).
    (d--f) Time evolution of the particle-density profiles for
    (d) the control, (e) attractive, and (f) repulsive cases.
    Symbols denote experimental measurements obtained from segmented
    fluorescence images, and solid curves denote the corresponding
    two-dimensional advection--diffusion simulations using the experimental
    parameters given in \S\ref{sec:Exp_results}.
    The velocity scale \(U_{\mathrm m}=70\,\mu\mathrm{m\,s^{-1}}\) was
    obtained from particle tracking near \(z=0\) in the control experiment.
    }
    \label{fig:Exp}
\end{figure*}

\subsection{\label{sec:Exp_setup}Experimental setup}

Experiments were conducted in a straight PDMS microchannel with a rectangular
cross-section, plasma-bonded to a glass coverslip
(figure~\ref{fig:Exp}a--c). We denote the streamwise, gapwise, and spanwise
coordinates by \(x\), \(y\), and \(z\), respectively, with \(y=0\) at the
gap centreline and \(z=0\) at the centre of the channel width. The channel had
width \(W=5\,\mathrm{mm}\), length \(L=25\,\mathrm{mm}\), and gap thickness
\(2a=25\,\mu\mathrm{m}\). Its large aspect ratio,
\(W/(2a)=200\), places the central portion of the channel in the Hele--Shaw
limit: sufficiently far from the lateral sidewalls, variations in the
spanwise direction \(z\) are negligible, whereas the dominant velocity
gradient occurs across the gap in the \(y\)-direction. The gap-resolved
\(x\)--\(y\) section near \(z=0\) can therefore be represented by the
two-dimensional plane-Poiseuille model of \S\ref{sec:Simulation}, which
retains the gapwise shear and cross-streamline particle migration of interest.

The channel was initially filled with fluorescent, negatively charged,
carboxylate-coated polystyrene colloids of diameter \(1\,\mu\mathrm{m}\),
suspended in an aqueous LiCl solution of concentration \(c_{\mathrm i}\).
At \(t=0\), the inlet was switched to a particle-free LiCl solution of
concentration \(c_{\mathrm f}\). The flow was driven by a syringe pump at a
constant volumetric flow rate. The mean velocity used in the two-dimensional
model was \(U_{\mathrm m}=70\,\mu\mathrm{m\,s^{-1}}\). Here, \(U_{\mathrm m}\) is the mean streamwise particle speed measured by
particle tracking near \(z=0\) in the control experiment. With no imposed
solute gradient, this measurement provides the gap-averaged pressure-driven
velocity used in the two-dimensional model.

Three displacement protocols were examined, corresponding to those in
\S\ref{sec:Simulation}. In the attractive case,
\(c_{\mathrm i}=0.1\,\mathrm{mM}\) and
\(c_{\mathrm f}=10.0\,\mathrm{mM}\), giving
\(\beta=c_{\mathrm f}/c_{\mathrm i}=100\). In the repulsive case,
\(c_{\mathrm i}=10.0\,\mathrm{mM}\) and
\(c_{\mathrm f}=0.1\,\mathrm{mM}\), giving \(\beta=0.01\). In the control
case, \(c_{\mathrm f}=c_{\mathrm i}\), so that \(\beta=1\).

Although the concentration ratio \(\beta\) is the only concentration
parameter entering the constant-mobility model, the absolute electrolyte
concentration can influence the quantitative particle response because the
Debye length, particle \(\zeta\)-potential, and hence
\(\Gamma_{\mathrm p}\) vary with ionic strength. The constant mobility used
in the simulations accompanying the experiments is based on previous
measurements for the same colloid--LiCl system
\citep{Li_Alipour_Pahlavan_2026}. Appendix~\ref{sec:cEffect} examines the
sensitivity to absolute concentration using a representative
finite-double-layer, concentration-dependent mobility model. Lowering the
concentration pair from \(0.1\)--\(10\,\mathrm{mM}\) to
\(0.001\)--\(0.1\,\mathrm{mM}\), while preserving the same concentration
ratio, modifies the magnitude of the particle redistribution but preserves
the qualitative front sharpening in the attractive case and front broadening
in the repulsive case.

Particle motion was recorded from above by fluorescence microscopy at
\(4\times\) magnification and a frame rate of \(10\,\mathrm{Hz}\). The field
of view was centred near \(z=0\), away from the lateral sidewalls. Each image
therefore recorded the particle distribution projected through the gap
direction \(y\). The images were processed using a custom MATLAB code. Each
frame was binarized to identify fluorescent particle regions, and their
locations were determined from the centroids of the segmented clusters.
The detections were then binned along \(x\) and averaged over the central
spanwise region to obtain the streamwise particle-density profiles.

\subsection{\label{sec:Exp_results}Comparison with simulations}

To compare with the experiments, we performed two-dimensional
advection--diffusion simulations using the experimental geometry and transport
parameters. The channel half-gap was set to \(a=12.5\,\mu\mathrm{m}\), the
salt diffusivity to
\(D_{\mathrm s}=1.37\times10^{-9}\,\mathrm{m}^2\,\mathrm{s}^{-1}\), and the
particle diffusivity to
\(D_{\mathrm p}=4.4\times10^{-13}\,\mathrm{m}^2\,\mathrm{s}^{-1}\), estimated
from the Stokes--Einstein relation for \(1\,\mu\mathrm{m}\) colloids. The
diffusiophoretic mobility was taken as
\(\Gamma_{\mathrm p}=0.7\times10^{-9}\,\mathrm{m}^2\,\mathrm{s}^{-1}\), based
on previous measurements for the same colloid--LiCl system
\citep{Li_Alipour_Pahlavan_2026}. The mean flow speed was set to the measured
value, \(U_{\mathrm m}=70\,\mu\mathrm{m\,s^{-1}}\). All other aspects of the
numerical setup follow \S\ref{sec:Simulation_setup}.

With these parameters,
\[
\tau_{\mathrm s}=\frac{a^2}{D_{\mathrm s}}\simeq0.11\,\mathrm{s},
\qquad
\tau_{\mathrm{PV}}=\frac{L}{U_{\mathrm m}}\simeq360\,\mathrm{s},
\qquad
\tau_{\mathrm p}=\frac{a^2}{D_{\mathrm p}}\simeq360\,\mathrm{s}.
\]
Thus the solute equilibrates across the gap on a timescale much shorter than
the residence time, whereas particle diffusion acts only over the residence
time itself. The experiments therefore probe the same Taylor-dispersed solute
regime as the simulations of \S\ref{sec:Simulation}, but with particle
diffusion expected to contribute more noticeably to the smoothing of sharp
particle features.

Figure~\ref{fig:Exp} compares the streamwise particle-density profiles from
experiments and simulations for the three protocols. The two diffusiophoretic
regimes are clearly distinguished. In the attractive case
(figure~\ref{fig:Exp}e), both experiment and simulation show a sharp, compact
leading edge with enhanced particle density near the front. In the repulsive
case (figure~\ref{fig:Exp}f), both show a broader trailing distribution,
consistent with migration of particles toward slower near-wall streamlines.
The control case (figure~\ref{fig:Exp}d) lies between these two limits and is
shaped primarily by the parabolic shear profile, as expected when no solute
gradient is imposed. The agreement is qualitative in all three cases and
semi-quantitative in front position and overall shape.

The remaining quantitative discrepancies, particularly in peak amplitudes, are
consistent with effects not included in the two-dimensional
advection--diffusion model. Diffusio-osmotic flows generated along the channel
walls can modify the local velocity profile and therefore the near-wall
transport that controls the trailing edge in the repulsive case. The
diffusiophoretic mobility was also treated as constant, whereas
\(\Gamma_{\mathrm p}\) can vary with local salt concentration. The effective
particle diffusivity is subject to uncertainty from particle polydispersity
and near-wall hydrodynamic effects. Finally, the fluorescence-based density
measurement relies on segmentation of particle clusters, which may
underestimate dense or overlapping regions, particularly near the leading edge
in the attractive case.

The central conclusion is that the qualitative signatures observed in the
simulations of \S\ref{sec:Simulation} are reproduced experimentally:
attractive fronts sharpen and accelerate particle removal, repulsive fronts
broaden and retain particles, and the control case is governed primarily by
shear. The cross-streamline migration mechanism is therefore a robust feature
of channel-flow diffusiophoresis rather than an artefact of the idealized
simulation geometry. We now turn to the analytical reduction that explains the
origin of these behaviours.

\section{\label{sec:Model}Non-diffusive particle transport model}

We now develop a reduced model to explain how a Taylor-dispersed solute front
can nevertheless drive cross-streamline particle migration. The model applies
in the intermediate regime
\[
  a^2/D_{\mathrm s}\ll t\ll a^2/D_{\mathrm p},
\]
in which the solute has equilibrated across the channel and entered the
Taylor-dispersion regime, while the colloids remain effectively non-diffusive
across the gap. Particles are then advected by the superposition of the
background Poiseuille flow and the diffusiophoretic drift generated by the
residual transverse solute gradients.

The development proceeds in three steps. We first derive the Taylor-regime
solute field, retaining the leading transverse correction
(\S\ref{subsec:Solute}). We then construct the effective particle velocity field
and identify approximate invariants of the resulting characteristic equations
(\S\ref{sec:Quasi_Stationary}--\S\ref{sec:Invariants}). For particles
initially far from the solute front,
\(|x_0|\gg D_{\mathrm{s,eff}}/U_{\mathrm m}\), where \(x_0\) is the initial streamwise position measured relative to the solute front and \(D_{\mathrm{s,eff}}\) the effective Taylor dispersion diffusivity, these invariants determine the
cross-streamline displacement and the density along characteristics. Finally,
we use the invariants to reconstruct the particle-occupied region in the
attractive (\S\ref{sec:Attractive}) and repulsive (\S\ref{sec:Repulsive})
cases, and validate the predictions against the simulations of
\S\ref{sec:Simulation} (\S\ref{subsec:Validation}).

\subsection{\label{subsec:Solute}Solute field in the Taylor-dispersion regime}

We first derive the solute field experienced by the particles after the solute
has diffused across the channel gap. The key point is that, although the
leading-order Taylor-dispersed solute concentration is cross-sectionally
uniform, the shear flow maintains a weak transverse correction. This correction
is small in the solute concentration but produces the transverse solute
gradient that drives cross-streamline diffusiophoretic migration.

The solute concentration in a two-dimensional pressure-driven channel of
half-width \(a\) satisfies
\begin{equation}
  \frac{\partial c}{\partial t}
  + u_x(y)\frac{\partial c}{\partial x}
  =
  D_{\mathrm s}
  \left(
  \frac{\partial^2 c}{\partial x^2}
  +
  \frac{\partial^2 c}{\partial y^2}
  \right),
  \qquad
  u_x(y)=\frac{3}{2}U_{\mathrm m}\left(1-\frac{y^2}{a^2}\right),
  \label{eq:solute_dimensional}
\end{equation}
with no-flux conditions \(\partial_y c=0\) at the centreline \(y=0\) and the
wall \(y=a\). The front is generated by displacing an initial solution of
concentration \(c_{\mathrm i}\) by an incoming solution of concentration
\(c_{\mathrm f}\). Defining
\begin{equation}
  \theta=\frac{c-c_{\mathrm i}}{c_{\mathrm f}-c_{\mathrm i}},
  \qquad
  \bar{y}=\frac{y}{a},
  \qquad
  \tilde x=\frac{x-U_{\mathrm m}t}{a},
  \qquad
  \tilde t=\frac{tD_{\mathrm s}}{a^2},
  \qquad
  \mathrm{Pe}_{\mathrm s}=\frac{U_{\mathrm m}a}{D_{\mathrm s}},
  \label{eq:Solute_ND_variables}
\end{equation}
the solute equation in the frame moving with the mean flow becomes
\begin{equation}
  \frac{\partial \theta}{\partial \tilde t}
  +
  \frac{1}{2}\mathrm{Pe}_{\mathrm s}(1-3\bar{y}^2)
  \frac{\partial \theta}{\partial \tilde x}
  =
  \frac{\partial^2\theta}{\partial \tilde x^2}
  +
  \frac{\partial^2\theta}{\partial \bar{y}^2}.
  \label{eq:solute_dimensionless}
\end{equation}

We seek the long-time Taylor-dispersion approximation for \(\tilde t\gg1\),
following the multi-scale derivation reviewed by \citet{camassa2010exact, linan2020taylor}.
Introducing the slow variables
\begin{equation}
  \tau=\epsilon \tilde t,
  \qquad
  \xi=\epsilon^{1/2}\tilde x,
  \qquad
  \epsilon\ll1,
\end{equation}
and expanding
\[
  \theta=\theta_0+\epsilon^{1/2}\theta_1+\epsilon\theta_2+\cdots,
\]
the half-power scaling of \(\xi\) relative to \(\tau\) is dictated by the
requirement that streamwise diffusion balance the leading shear-induced
transverse correction at the same order. At successive orders, we obtain
\begin{align}
  \mathcal O(1):\qquad
  &\partial_{\bar{y}}^2\theta_0=0,
  \label{eq:solute_order0}
  \\
  \mathcal O(\epsilon^{1/2}):\qquad
  &\partial_{\bar{y}}^2\theta_1
  =
  \frac{1}{2}\mathrm{Pe}_{\mathrm s}(1-3\bar{y}^2)
  \partial_{\xi}\theta_0,
  \label{eq:solute_order1}
  \\
  \mathcal O(\epsilon):\qquad
  &\partial_{\bar{y}}^2\theta_2
  =
  \partial_{\tau}\theta_0
  +
  \frac{1}{2}\mathrm{Pe}_{\mathrm s}(1-3\bar{y}^2)
  \partial_{\xi}\theta_1
  -
  \partial_{\xi}^2\theta_0 .
  \label{eq:solute_order2}
\end{align}

The leading-order solution is independent of \(\bar{y}\):
\(\theta_0=\theta_0(\xi,\tau)\).

Solving \eqref{eq:solute_order1} subject to the no-flux conditions gives
\begin{equation}
  \theta_1
  =
  \frac{\mathrm{Pe}_{\mathrm s}}{8}
  \left(
  \bar{y}^2(2-\bar{y}^2)
  \right)
  \frac{\partial \theta_0}{\partial \xi} 
  + \theta_{\text{1a}}(\xi,\tau),
  \label{eq:c1_solution}
\end{equation}
where \(\theta_{\text{1a}}(\xi,\tau)\) depends only on \(\xi\) and \(\tau\) so it does not contribute to the transverse solute gradient.

Integrating \eqref{eq:solute_order1} over the half-channel and noting the no-flux boundary condition \(\partial_{\bar y}\theta_2|_{\bar y = 0,1}=0\) gives
\begin{equation}
\int_0^1
\left[
\partial_{\tau}\theta_0
+
\frac{1}{2}\mathrm{Pe}_{\mathrm s}(1-3{\bar{y}}^2)\partial_{\xi}\theta_1
-
\partial_{\xi}^2\theta_0
\right]d{\bar{y}}
=0 .
\label{eq:Solvability_Condition_general}
\end{equation}
Substituting \eqref{eq:c1_solution}, and using
\[
  \int_0^1(1-3{\bar{y}}^2){\bar{y}}^2(2-{\bar{y}}^2)\,d{\bar{y}}=-\frac{32}{105},
\]
gives the Taylor-dispersion equation for the leading-order field,
\begin{equation}
  \frac{\partial \theta_0}{\partial \tau}
  =
  \bar D_{\mathrm{s,eff}}
  \frac{\partial^2\theta_0}{\partial \xi^2},
  \qquad
  \bar D_{\mathrm{s,eff}}
  =
  1+\frac{2}{105}\mathrm{Pe}_{\mathrm s}^2 .
  \label{eq:taylor_dispersion_c0}
\end{equation}

For an initially sharp solute step, the leading-order front in dimensional
variables is
\begin{equation}
  X(x,t)
  =
  \frac{1}{2}
  \left[
  1-\mathrm{erf}
  \left(
  \frac{x-U_{\mathrm m}t}{\sqrt{4D_{\mathrm{s,eff}}t}}
  \right)
  \right],
  \qquad
  D_{\mathrm{s,eff}}
  =
  D_{\mathrm s}
  \left(
  1+\frac{2}{105}\mathrm{Pe}_{\mathrm s}^2
  \right).
  \label{eq:Xfunction}
\end{equation}
Combining the leading-order concentration with the first transverse correction
gives
\begin{equation}
  c(x,y,t)
  =
  c_{\mathrm i}
  +(c_{\mathrm f}-c_{\mathrm i})
  \left[
  X(x,t)
  +
  \frac{\mathrm{Pe}_{\mathrm s}}{8}
  \left(\frac{y}{a}\right)^2
  \left[
  2-\left(\frac{y}{a}\right)^2
  \right]
  a\,\frac{\partial X}{\partial x}
  + \sqrt{\epsilon} \theta_{\text{1a}} \right]
  +\mathcal O(\epsilon).
  \label{eq:soluteFieldExpr}
\end{equation}
Differentiating \eqref{eq:soluteFieldExpr} in \(y\) gives the central result of
this subsection: the residual transverse solute gradient
\begin{equation}
  \frac{\partial c}{\partial y}
  =
  (c_{\mathrm f}-c_{\mathrm i})
  \frac{\mathrm{Pe}_{\mathrm s}}{2}
  \frac{y}{a}
  \left[
  1-\left(\frac{y}{a}\right)^2
  \right]
  \frac{\partial X}{\partial x}
  +\mathcal O(\epsilon).
  \label{eq:solute_transverse_gradient}
\end{equation}
This gradient vanishes at the centreline and wall but is finite in the channel
interior. Two features of \eqref{eq:solute_transverse_gradient} are essential
for what follows. First, the prefactor \(\mathrm{Pe}_{\mathrm s}\) shows that
the transverse gradient is enhanced relative to the axial gradient by the
solute P\'eclet number. Second, because
\(\partial_x X\sim 1/\sqrt{D_{\mathrm{s,eff}}t}\) for a step initial condition,
the transverse gradient decays only as \(t^{-1/2}\), on the slow
Taylor-dispersion timescale of the solute front rather than on the fast
cross-channel diffusion timescale \(\tau_{\mathrm s}\).

\subsection{\label{sec:Quasi_Stationary}Effective particle dynamics}

We now use the Taylor-regime solute field \(c(x,y,t)\) to construct the
effective particle velocity field \(\mathbf V\). Denoting the background flow
velocity by \(\mathbf u\) and the particle diffusiophoretic mobility by
\(\Gamma_{\mathrm p}\), we write
\begin{equation}
  \mathbf V=\mathbf u+\Gamma_{\mathrm p}\nabla\ln c,
  \label{eq:Veff}
\end{equation}
since particle diffusion is subdominant
\((\tau_{\mathrm p}\gg\tau_{\mathrm{PV}})\) and so does not contribute to the
leading-order transport, although it smooths the sharp interfaces predicted by
the non-diffusive dynamics, as discussed in \S\ref{subsec:Validation}. We define the solute concentration ratio \(\beta\) and the corresponding
leading-order dimensionless concentration \(C_\beta\) as
\begin{equation}
  \beta=\frac{c_{\mathrm f}}{c_{\mathrm i}},
  \qquad
  C_\beta(x,t)=1+(\beta-1)X(x,t).
  \label{eq:Cbeta_dimensional}
\end{equation}
The leading-order solute concentration is then
\(c\approx c_{\mathrm i}C_\beta\). Throughout, we take
\(\Gamma_{\mathrm p}>0\), so the sign of the cross-streamline migration is
determined by \(\beta-1\).

The streamwise component of \eqref{eq:Veff} contains both the Poiseuille
profile and a streamwise diffusiophoretic correction. The latter is small in
the regime of interest:
\begin{equation}
  \frac{U_{\mathrm{DP},x}}{U_{\mathrm m}}
  \sim
  \frac{\Gamma_{\mathrm p}|\ln\beta|}
  {U_{\mathrm m}\sqrt{D_{\mathrm{s,eff}}t}}
  \ll 1
  \qquad \text{for} \qquad
  t\gg
  \frac{\Gamma_{\mathrm p}^2\ln^2\beta}
  {U_{\mathrm m}^2 D_{\mathrm{s,eff}}}.
  \label{eq:streamwise_DP_scaling}
\end{equation}
For the parameters of \S\ref{sec:Simulation}, this threshold is
\(\simeq0.7\,\mathrm{s}\), much shorter than the residence time
\(\tau_{\mathrm{PV}}=300\,\mathrm{s}\). Similarly, in evaluating the
transverse component \(V_y=\Gamma_{\mathrm p}\partial_y\ln c\), retaining only
the cross-sectionally uniform leading-order solute concentration in the
denominator introduces an error of order
\[
  \mathrm{Pe}_{\mathrm s}a|\partial_x X|
  \sim
  \left[\frac{t}{\tau_{\mathrm s}^2/\tau_{\mathrm c}}\right]^{-1/2},
  \qquad
  \tau_{\mathrm c}=\frac{D_{\mathrm{s,eff}}}{U_{\mathrm m}^2}.
\]
For the same parameters, this correction becomes small after
\(\simeq1.6\,\mathrm{s}\). Both approximations therefore apply over essentially
the entire residence time. The effective velocity components reduce to
\begin{subequations}
\label{eq:Ueff_dim}
\begin{align}
V_x
&\approx
\frac{3}{2}U_{\mathrm m}\left(1-\frac{y^2}{a^2}\right),
\label{eq:Veff_x}
\\
V_y
&\approx
\frac{1}{2}\Gamma_{\mathrm p}\,\mathrm{Pe}_{\mathrm s}\,
\frac{y}{a}\left[1-\left(\frac{y}{a}\right)^2\right]
\partial_x\ln C_\beta .
\label{eq:Veff_y}
\end{align}
\end{subequations}
The transverse velocity carries an explicit factor of
\(\mathrm{Pe}_{\mathrm s}\): the residual transverse solute gradient is
P\'eclet-enhanced relative to the axial gradient, and this enhancement is
inherited by the diffusiophoretic drift.

We nondimensionalize the particle problem using the Taylor-dispersion length
and time scales,
\[
  L_x\equiv U_{\mathrm m}\tau_{\mathrm c}=\frac{D_{\mathrm{s,eff}}}{U_{\mathrm m},
  }
  \qquad
  \tau_{\mathrm c}=\frac{D_{\mathrm{s,eff}}}{U_{\mathrm m}^2}.
\]
Equivalently,
\begin{equation}
  \bar x=\frac{x-U_{\mathrm m}t}{U_{\mathrm m}\tau_{\mathrm c}},
  \qquad
  \bar y=\frac{y}{a},
  \qquad
  \bar t=\frac{t}{\tau_{\mathrm c}},
  \qquad
  \bar\Gamma_{\mathrm p}=\frac{\Gamma_{\mathrm p}}{D_{\mathrm s}} .
  \label{eq:particle_nondim}
\end{equation}
These scales are chosen so that the solute front has \(O(1)\) width in the
dimensionless streamwise coordinate when \(\bar t=O(1)\). In these variables,
\begin{equation}
  \bar X(\bar x,\bar t)
  =
  \frac12
  \left[
  1-\erf\left(\frac{\bar x}{\sqrt{4\bar t}}\right)
  \right],
  \qquad
  C_\beta(\bar x,\bar t)
  =
  1+(\beta-1)\bar X(\bar x,\bar t).
  \label{eq:XandC}
\end{equation}
For the streamwise velocity, the natural rescaling is
\(\bar V_x=V_x/U_{\mathrm m}-1\), since we work in the frame moving with the
mean flow. For the transverse velocity, the consistent scaling is
\(\bar V_y=(\tau_{\mathrm c}/a)V_y\). Using
\(\partial_x=(U_{\mathrm m}\tau_{\mathrm c})^{-1}\partial_{\bar x}\) and
\(\mathrm{Pe}_{\mathrm s}=U_{\mathrm m}a/D_{\mathrm s}\), we obtain
\begin{equation}
  \bar V_x
  =
  \frac12(1-3\bar y^2),
  \qquad
  \bar V_y
  =
  \frac12\bar\Gamma_{\mathrm p}\,
  \bar y(1-\bar y^2)\,
  \partial_{\bar x}\ln C_\beta .
  \label{eq:Ueff_xy}
\end{equation}
The factor \(\mathrm{Pe}_{\mathrm s}\) present in the dimensional \(V_y\) is
absorbed by the Taylor-dispersion length and time scales, leaving
\(\bar\Gamma_{\mathrm p}=\Gamma_{\mathrm p}/D_{\mathrm s}\) as the
dimensionless measure of transverse phoretic migration. In these scaled
variables, the reduced characteristic dynamics depend only on
\(\bar\Gamma_{\mathrm p}\) and \(\beta\); the dependence on
\(\mathrm{Pe}_{\mathrm s}\) enters through the dimensional Taylor-dispersion
scales and through the validity of the asymptotic reduction. Equation
\eqref{eq:Ueff_xy} is the central reduced description of the particle dynamics.

\subsection{\label{sec:Invariants}Characteristic invariants and density along trajectories}

The non-diffusive particle equation reads
\begin{equation}
  \partial_{\bar t} n+\bar\nabla\cdot(n\bar{\mathbf V})=0,
  \label{eq:n_advection}
\end{equation}
where \(\bar{\mathbf V} = (\bar V_x, \bar V_y)\), or, equivalently for \(\phi=\ln n\),
\begin{equation}
  \partial_{\bar t}\phi
  +\bar V_x\partial_{\bar x}\phi
  +\bar V_y\partial_{\bar y}\phi
  =
  -\partial_{\bar y}\bar V_y .
  \label{eq:moc_pde}
\end{equation}
Along particle characteristics, this gives
\begin{align}
  \dot{\bar x}
  &=
  \frac12(1-3\bar y^2),
  \label{eq:char_x}
  \\
  \dot{\bar y}
  &=
  \frac12\bar\Gamma_{\mathrm p}
  \bar y(1-\bar y^2)
  \partial_{\bar x}\ln C_\beta,
  \label{eq:char_y}
  \\
  \dot{\phi}
  &=
  -\frac12\bar\Gamma_{\mathrm p}
  (1-3\bar y^2)
  \partial_{\bar x}\ln C_\beta .
  \label{eq:char_phi}
\end{align}
We now show that \eqref{eq:char_x}--\eqref{eq:char_phi} admit two approximate
invariants: one constraining the cross-streamline displacement, and one giving
the particle density along characteristics. These invariants are the basis of
the analytical reconstruction in the remaining subsections.

Define
\begin{equation}
  F(\bar y)
  =
  \frac{3\sqrt3}{2}\bar y(1-\bar y^2),
  \label{eq:F_def}
\end{equation}
so that \(0\leq F\leq1\) for \(0\leq\bar y\leq1\), with maximum at the
mean-flow streamline
\[
  \bar y_{\mathrm c}=\frac{1}{\sqrt3}.
\]
Since
\[
  F'(\bar y)
  =
  \frac{3\sqrt3}{2}(1-3\bar y^2)
  =
  3\sqrt3\,\bar V_x(\bar y),
\]
we have
\begin{equation}
  \frac{d}{d\bar t}\ln F(\bar y)
  =
  \frac{F'(\bar y)}{F(\bar y)}\dot{\bar y}
  =
  \bar\Gamma_{\mathrm p}\bar V_x
  \partial_{\bar x}\ln C_\beta .
  \label{eq:dlogFdt}
\end{equation}
Similarly, \eqref{eq:char_phi} gives
\begin{equation}
  \frac{d\phi}{d\bar t}
  =
  -\bar\Gamma_{\mathrm p}\bar V_x
  \partial_{\bar x}\ln C_\beta .
  \label{eq:dphidt}
\end{equation}
The two right-hand sides are equal and opposite. Using the chain rule along a
characteristic,
\begin{equation}
  \bar V_x\partial_{\bar x}\ln C_\beta
  =
  \frac{d}{d\bar t}\ln C_\beta
  -
  \partial_{\bar t}\ln C_\beta,
  \label{eq:dlogCdt}
\end{equation}
and integrating from \(0\) to \(\bar t\) gives
\begin{align}
  \ln F(\bar y)
  -\bar\Gamma_{\mathrm p}\ln C_\beta(\bar x,\bar t)
  &=
  \ln F(\bar y_0)
  -\bar\Gamma_{\mathrm p}\ln C_\beta(\bar x_0,0)
  -
  \bar\Gamma_{\mathrm p}\,\mathcal I(\bar x_0,\bar t),
  \label{eq:invariant_y_full}
  \\
  \phi
  +\bar\Gamma_{\mathrm p}\ln C_\beta(\bar x,\bar t)
  &=
  \phi_0
  +\bar\Gamma_{\mathrm p}\ln C_\beta(\bar x_0,0)
  +
  \bar\Gamma_{\mathrm p}\,\mathcal I(\bar x_0,\bar t),
  \label{eq:invariant_phi_full}
\end{align}
where
\begin{equation}
  \mathcal I(\bar x_0,\bar t)
  =
  \int_0^{\bar t}
  \partial_s\ln C_\beta(\bar x(s),s)\,ds .
  \label{eq:I_def}
\end{equation}
Adding \eqref{eq:invariant_y_full} and
\eqref{eq:invariant_phi_full} eliminates both the concentration-dependent
terms and the remainder \(\mathcal I\), yielding the exact invariant
\begin{equation}
    n(\bar x,\bar y,\bar t)F(\bar y)
    =
    n(\bar x_0,\bar y_0,0)F(\bar y_0).
    \label{eq:invariant_nF}
\end{equation}
Thus, the local particle density and transverse position are coupled exactly
along each characteristic: changes in \(F(\bar y)\) are accompanied by
reciprocal changes in \(n\).

To obtain separate invariants for the transverse position and particle
density, we now estimate the remainder \(\mathcal I\).
Physically,
\(\mathcal I\) measures the cumulative effect of the slow temporal broadening
of the solute front along a particle trajectory. For
\(|\bar x_0|\gg 1\), the particle interacts with the front over a time
interval that is short compared with the timescale on which the front
broadens appreciably, and the accumulated contribution is therefore small.
As shown in Appendix~\ref{app:integral},
\(\mathcal I=O(|\bar x_0|^{-1/2})\) as \(|\bar x_0|\to\infty\).
Neglecting this remainder gives the approximate invariants
\begin{align}
  F(\bar y)C_\beta(\bar x,\bar t)^{-\bar\Gamma_{\mathrm p}}
  &=
  F(\bar y_0)C_\beta(\bar x_0,0)^{-\bar\Gamma_{\mathrm p}},
  \label{eq:invariant_y}
  \\
  n(\bar x,\bar y,\bar t)
  C_\beta(\bar x,\bar t)^{\bar\Gamma_{\mathrm p}}
  &=
  n(\bar x_0,\bar y_0,0)
  C_\beta(\bar x_0,0)^{\bar\Gamma_{\mathrm p}}.
  \label{eq:invariant_n}
\end{align}
Equation~\eqref{eq:invariant_y} constrains the transverse positions that can
be reached from a given initial streamline, whereas
\eqref{eq:invariant_n} determines how the particle density varies with the
local solute concentration along a trajectory.

The initial particle distribution lies downstream of the solute step, so that
\(\bar x_0>0\) and \(\bar X(\bar x_0,0)=0\), giving
\(C_\beta(\bar x_0,0)=1\). From \eqref{eq:invariant_n}, the particle density along any reachable
characteristic is therefore
\begin{equation}
  n(\bar x,\bar y,\bar t)
  =
  n_0 C_\beta(\bar x,\bar t)^{-\bar\Gamma_{\mathrm p}},
  \label{eq:n_positiveX0}
\end{equation}
or, written explicitly in terms of \(\bar x\) and \(\bar t\),
\begin{equation}
  n(\bar x,\bar y,\bar t)
  =
  n_0
  \left[
  \frac{\beta+1}{2}
  -
  \frac{\beta-1}{2}
  \erf\left(\frac{\bar x}{\sqrt{4\bar t}}\right)
  \right]^{-\bar\Gamma_{\mathrm p}},
  \label{eq:n_positiveX0_explicit}
\end{equation}
where we assume a uniform initial condition \(n(\bar{x}_0,\bar{y}_0,0) = n_0\). 
Equation~\eqref{eq:n_positiveX0_explicit} gives the density only in regions
reached by particle characteristics. Regions not reached remain particle-free.

The corresponding cross-streamline displacement is determined by the invariant
\begin{equation}
  F(\bar y) C_\beta(\bar x,\bar t)^{-\bar\Gamma_{\mathrm p}}
  =
  F(\bar y_0).
  \label{eq:surface_positiveX0}
\end{equation}
Particles released from a given initial streamline \(\bar y_0\) at different
downstream positions \(\bar x_0>0\) all lie on the same curve in
\((\bar x,\bar y)\) at any given time. As \(\bar y_0\) varies, these curves
sweep out the region occupied by particles. The particle field is therefore
determined by a reachability problem: \(n\) is given by
\eqref{eq:n_positiveX0_explicit} inside the region swept out by characteristics
and is zero outside it.

In the next two subsections, we use \eqref{eq:surface_positiveX0} together with
the Poiseuille envelope to approximate the moving particle interface, namely
the boundary of the particle-occupied region in \((\bar x,\bar y)\). The
attractive case \((\beta>1)\) and the repulsive case \((0<\beta<1)\) lead to
qualitatively different geometries.

\subsection{\label{sec:Attractive}Attractive case}

We first consider the attractive case,
\(\beta>1\) \((c_{\mathrm f}>c_{\mathrm i})\). For positive diffusiophoretic
mobility, particles migrate up the solute gradient. Within the Taylor-regime
solute front this drives a transverse velocity directed toward the centreline,
transferring particles to faster streamlines.

It is useful to begin with the front that would be obtained in the absence of
diffusiophoresis. In the moving frame, particles on the streamline \(\bar y\)
travel with velocity \(\bar V_x=(1-3\bar y^2)/2\), so an initial interface
located at \(\bar x_0=0\) is sheared into the Poiseuille envelope
\begin{equation}
  \bar Y_{\mathrm F}(\bar x,\bar t)
  =
  \sqrt{\frac{1-2\bar x/\bar t}{3}},
  \qquad
  -\bar t\leq \bar x\leq \frac{\bar t}{2}.
  \label{eq:YF}
\end{equation}
The mean-flow streamline \(\bar y_{\mathrm c}=1/\sqrt3\) is stationary in this
frame. Particles with \(\bar y<\bar y_{\mathrm c}\) move ahead of the
solute-front centre, whereas those with \(\bar y>\bar y_{\mathrm c}\) lag
behind it. The interface near and behind the front is therefore controlled by
particles initially released on the slower streamlines,
\(\bar y_0\geq\bar y_{\mathrm c}\).

For \(\beta>1\), the leading-order dimensionless concentration
\begin{equation}
  C_\beta(\bar x,\bar t)
  =
  \frac{\beta+1}{2}
  -
  \frac{\beta-1}{2}
  \erf\left(\frac{\bar x}{\sqrt{4\bar t}}\right)
\end{equation}
approaches \(1\) as \(\bar x\to+\infty\) and \(\beta\) as
\(\bar x\to-\infty\). Within the approximate-invariant description, the diffusiophoretic
boundary of the particle-occupied region is determined by the limiting
wall-side characteristic that approaches the critical streamline
\(\bar y_{\mathrm c}=1/\sqrt3\) only as
\(\bar x\to-\infty\). Since \(F(\bar y_{\mathrm c})=1\), equation \eqref{eq:surface_positiveX0} fixes the invariant level of this
separatrix as
\begin{equation}
  F(\bar y)
  C_\beta(\bar x,\bar t)^{-\bar\Gamma_{\mathrm p}}
  =
  F(\bar y_0)
  =
  F(\bar y_{\mathrm c})
  C_\beta(-\infty,\bar t)^{-\bar\Gamma_{\mathrm p}}
  =
  \beta^{-\bar\Gamma_{\mathrm p}}.
  \label{eq:A_Invariant_y}
\end{equation}
The initial separatrix label therefore satisfies
\(F(\bar y_0)=\beta^{-\bar\Gamma_{\mathrm p}}\), which has two
roots in the physical interval,
\(\bar y_{\mathrm a2}<\bar y_{\mathrm c}<\bar y_{\mathrm a1}\):
\begin{equation}
  \bar y_{\mathrm a1}
  =
  \frac{2}{\sqrt3}
  \cos\!\left[
  \frac{\pi-\cos^{-1}(\beta^{-\bar\Gamma_{\mathrm p}})}{3}
  \right],
  \qquad
  \bar y_{\mathrm a2}
  =
  -\frac{2}{\sqrt3}
  \cos\!\left[
  \frac{2\pi-\cos^{-1}(\beta^{-\bar\Gamma_{\mathrm p}})}{3}
  \right].
  \label{eq:ya_roots}
\end{equation}
Only \(\bar y_{\mathrm a1}\) lies on the upper branch
\(\bar y_0>\bar y_{\mathrm c}\) and is therefore the initial label of
the separatrix. The second root, \(\bar y_{\mathrm a2}\), is on the lower branch of the same separatrix.

Similarly, solving \eqref{eq:A_Invariant_y} for \(\bar{y}\) gives both the upper and lower branches of the separatrix. The lower branch with
\(\bar y\leq\bar y_{\mathrm c}\) defines the diffusiophoretic envelope of the
interface
\begin{equation}
  \bar Y_{\mathrm{DP,A}}(\bar x,\bar t)
  =
  -\frac{2}{\sqrt3}
  \cos\!\left[
  \frac{2\pi-\cos^{-1}C_{\mathrm A}(\bar x,\bar t)}{3}
  \right],
  \qquad
  C_{\mathrm A}(\bar x,\bar t)
  =
  \left[
  \frac{C_\beta(\bar x,\bar t)}{\beta}
  \right]^{\bar\Gamma_{\mathrm p}} .
  \label{eq:YDP_A}
\end{equation}

\begin{figure*}
    \centering
    \includegraphics[width=\textwidth,keepaspectratio]{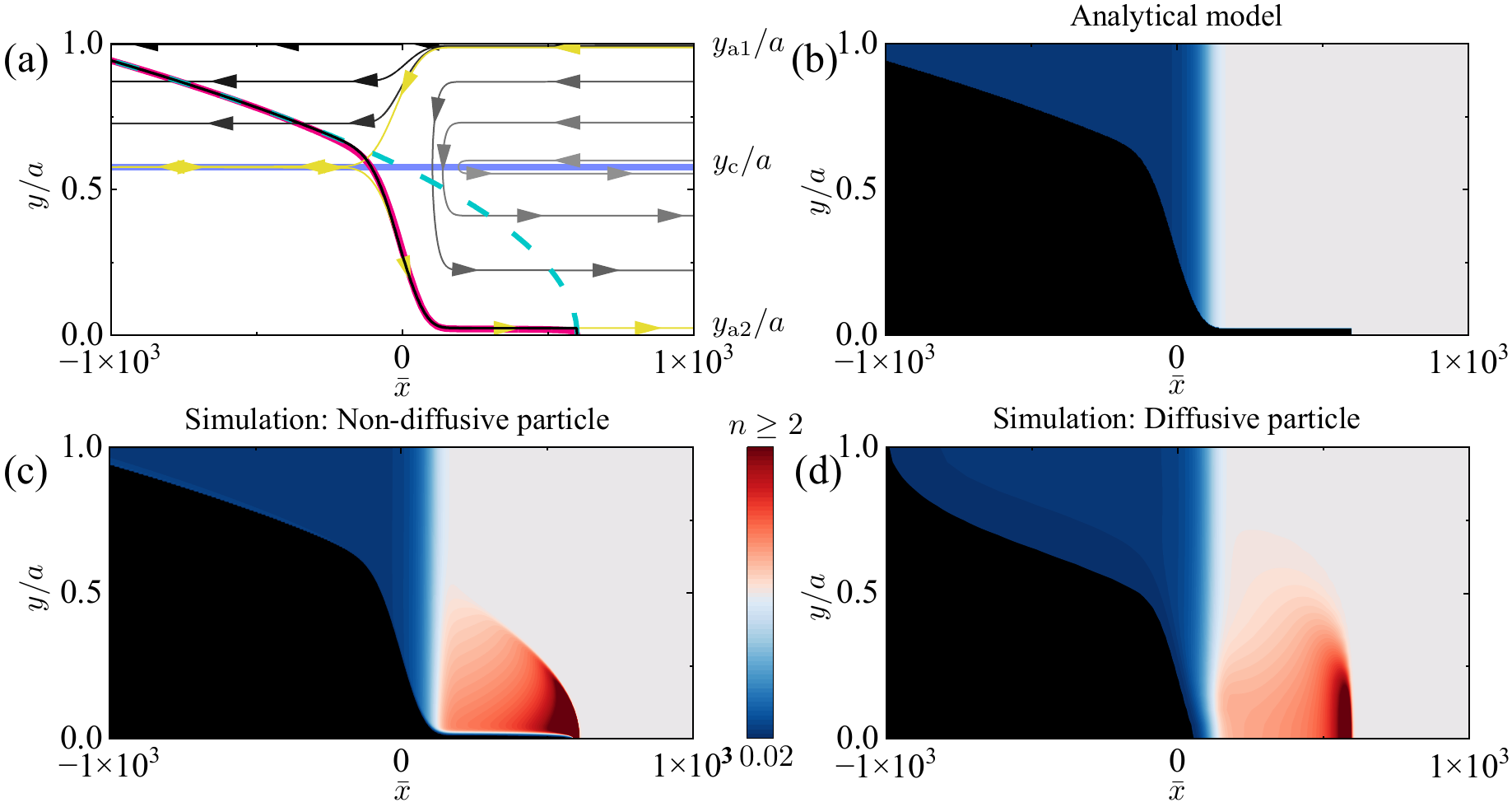}
    \captionsetup{width=\textwidth,justification=justified,singlelinecheck=false}
\caption{
Attractive case, \(\beta>1\), at \(\bar t=1200\)
(\(t/\tau_{\mathrm{PV}}\simeq0.43\) for the parameters of
\S\ref{sec:Simulation}).
(a) Interface geometry in the \((\bar x,\bar y)\) plane. The magenta curve is
the numerically tracked interface obtained by advecting the initial line
\(\bar x_0=0\) with the full effective velocity field. The cyan dashed curve
is the Poiseuille envelope \(\bar Y_{\mathrm F}\), and the thin solid curves
are representative constraint curves from \eqref{eq:surface_positiveX0},
labelled by their initial streamline \(\bar y_0\geq\bar y_{\mathrm c}\). The
yellow curve is the separatrix defining the diffusiophoretic envelope
\(\bar Y_{\mathrm{DP,A}}\), and the black curve is the composite approximation
\(\bar Y_{\mathrm A}\).
(b) Analytical reconstruction \(n_{\mathrm A}\).
(c) Pure-advection simulation of \eqref{eq:n_advection}.
(d) Full advection--diffusion simulation of \eqref{eq:n_ade}.
All density panels use the same colour scale, with saturation for \(n\geq2\).
}
    \label{fig:Attractive}
\end{figure*}

Figure~\ref{fig:Attractive}(a) compares this diffusiophoretic envelope with
the numerically tracked interface obtained by advecting the initial line
\(\bar x_0=0\) using the full effective velocity field. The tracked interface
follows the Poiseuille envelope far from the solute front, where
diffusiophoretic migration is weak, and the diffusiophoretic separatrix near
the front, where transverse migration is strongest. We therefore approximate
the interface by a geometric closure that combines the two:
\begin{equation}
\bar Y_{\mathrm A}(\bar x,\bar t)=
\begin{cases}
\bar Y_{\mathrm F}(\bar x,\bar t)\,
\dfrac{\bar Y_{\mathrm{DP,A}}(\bar x,\bar t)}
      {\bar Y_{\mathrm{DP,A}}(-\infty,\bar t)},
& -\bar t\leq \bar x\leq 0,\\[8pt]
\min\!\left[
\bar Y_{\mathrm F}(\bar x,\bar t),
\bar Y_{\mathrm{DP,A}}(\bar x,\bar t)
\right],
& 0<\bar x\leq \bar t/2 .
\end{cases}
\label{eq:YA}
\end{equation}
Behind the front centre \((\bar x\leq0)\), the interface is set primarily by
the diffusiophoretic separatrix, with the multiplicative Poiseuille factor
enforcing the wall constraint at \(\bar x=-\bar t\). Ahead of the front centre
\((0<\bar x\leq\bar t/2)\), the interface is set by whichever of the two
envelopes lies lower. Equation~\eqref{eq:YA} is not an exact characteristic
solution; rather, it is the simplest closure that reproduces the limiting
behaviour observed in the tracked interface.

The particle density inside the region reached by characteristics follows
from \eqref{eq:n_positiveX0}. Defining
\begin{equation}
g_{\mathrm A}(\bar x,\bar t)
=
n_0
\left[
\frac{\beta+1}{2}
-
\frac{\beta-1}{2}
\erf\left(\frac{\bar x}{\sqrt{4\bar t}}\right)
\right]^{-\bar\Gamma_{\mathrm p}},
\label{eq:g_A}
\end{equation}
the attractive-case particle field is
\begin{equation}
n_{\mathrm A}(\bar x,\bar y,\bar t)=
\begin{cases}
g_{\mathrm A}(\bar x,\bar t),
& \bar x>\bar t/2,\quad 0\leq\bar y\leq1,\\[3pt]
g_{\mathrm A}(\bar x,\bar t),
& -\bar t\leq\bar x\leq\bar t/2,\quad
  \bar y\geq\bar Y_{\mathrm A}(\bar x,\bar t),\\[3pt]
0,
& \text{otherwise}.
\end{cases}
\label{eq:nField_xy_A}
\end{equation}

Figures~\ref{fig:Attractive}(b--d) show that this reconstruction captures the
dominant structure of the attractive front: particles occupy a region bounded
below by an interface pulled toward the centreline, and the density is
amplified according to the local solute contrast. The reconstruction does not
capture the narrow, high-density plume near the centreline ahead of the front
seen in the simulations. That plume is generated by particles initially close
to the solute interface, for which the integral remainder
\(\mathcal I(\bar x_0,\bar t)\) is not small and the simple invariants
\eqref{eq:invariant_y}--\eqref{eq:invariant_n} break down. Particle diffusion
in the full advection--diffusion simulation further smooths the sharp
interface predicted by the non-diffusive reconstruction.

\subsection{\label{sec:Repulsive}Repulsive case}

We next consider the repulsive case,
\(0<\beta<1\) \((c_{\mathrm f}<c_{\mathrm i})\). For positive diffusiophoretic
mobility, particles still migrate up the solute gradient and therefore away
from the low-concentration invading fluid. Within the Taylor-regime solute
front this drives a transverse velocity directed toward the wall, pushing
particles toward slower streamlines and enhancing the near-wall trailing
region.

For \(0<\beta<1\), the leading-order dimensionless concentration
\begin{equation}
  C_\beta(\bar x,\bar t)
  =
  \frac{\beta+1}{2}
  -
  \frac{\beta-1}{2}
  \erf\left(\frac{\bar x}{\sqrt{4\bar t}}\right)
  \label{eq:Cbeta_R}
\end{equation}
increases monotonically from \(C_\beta=\beta\) far behind the front to
\(C_\beta=1\) far ahead of the front. The characteristic constraint is again
\begin{equation}
F(\bar y)
  C_\beta(\bar x,\bar t)^{-\bar\Gamma_{\mathrm p}}
  =
  F(\bar y_0).
  \label{eq:repulsive_constraint}
\end{equation}

As before, the streamline \(\bar y_{\mathrm c}=1/\sqrt3\) is stationary in the
moving frame, and the interface near the solute front is controlled by
particles initially released on the slower streamlines,
\(\bar y_0\geq\bar y_{\mathrm c}\). For the repulsive case, however,
diffusiophoresis bends these characteristic curves \emph{upward}, away from
the centreline. The lower boundary of the occupied near-wall region is
therefore generated by the smallest such label,
\(\bar y_0=\bar y_{\mathrm c}\). Since \(F(\bar y_{\mathrm c})=1\), this
boundary satisfies
\[
  F(\bar Y_{\mathrm{DP,R}})
  =
  C_\beta(\bar x,\bar t)^{\bar\Gamma_{\mathrm p}} .
\]
Because particles in the repulsive case are pushed toward the wall, we take
the upper branch of \(F^{-1}\), \(\bar Y_{\mathrm{DP,R}}\geq\bar y_{\mathrm c}\):
\begin{equation}
  \bar Y_{\mathrm{DP,R}}(\bar x,\bar t)
  =
  \frac{2}{\sqrt3}
  \cos\!\left[
  \frac{\pi-\cos^{-1}C_{\mathrm R}(\bar x,\bar t)}{3}
  \right],
  \qquad
  C_{\mathrm R}(\bar x,\bar t)
  =
  C_\beta(\bar x,\bar t)^{\bar\Gamma_{\mathrm p}}.
  \label{eq:YDP_R}
\end{equation}
This diffusiophoretic envelope approaches \(\bar y_{\mathrm c}\) as
\(\bar x\to+\infty\), where particles have not yet encountered the front, and
is displaced toward the wall behind the front.

\begin{figure*}
    \centering
    \includegraphics[width=\textwidth,keepaspectratio]{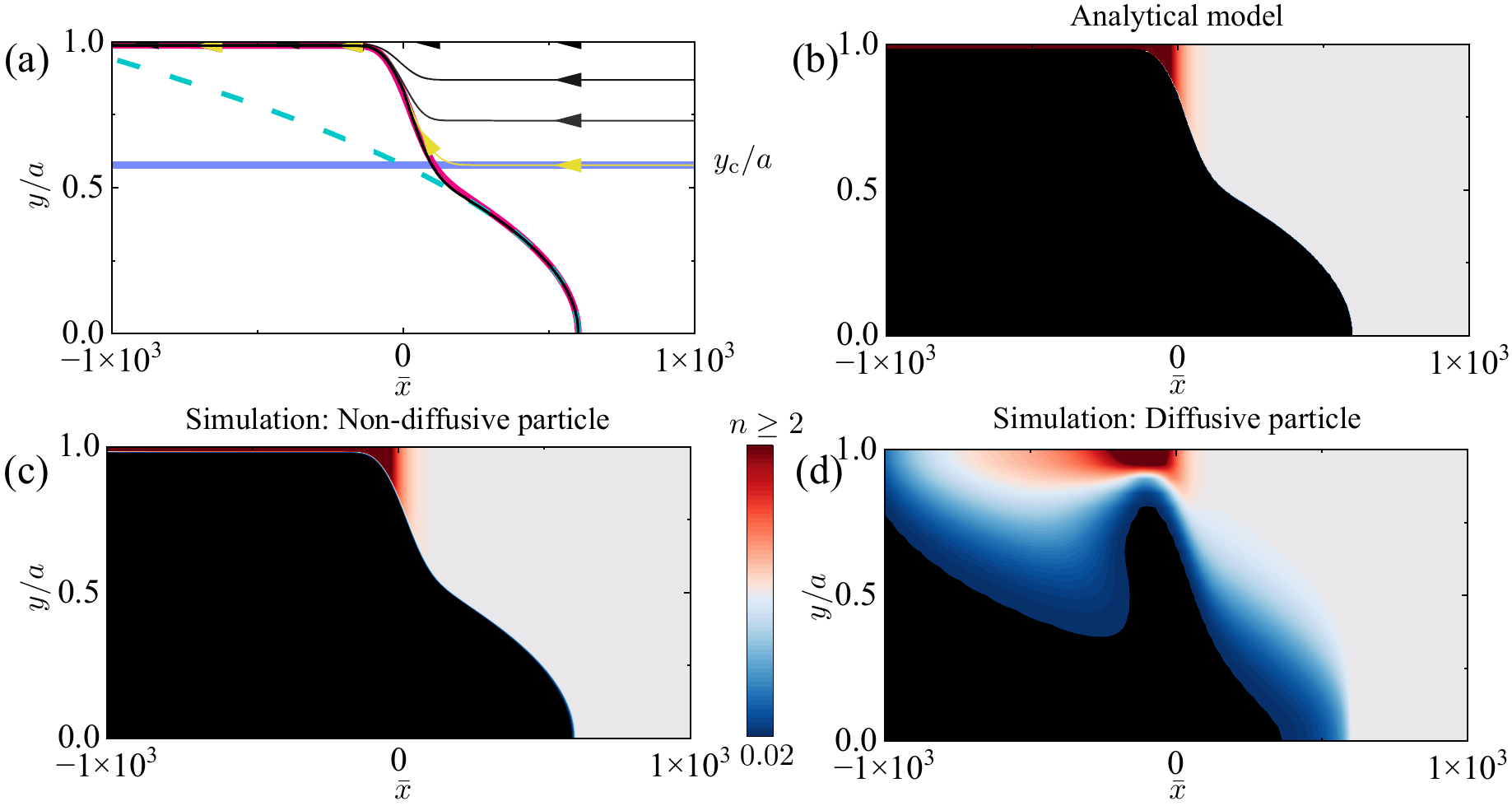}
    \captionsetup{width=\textwidth,justification=justified,singlelinecheck=false}
    \caption{
    Repulsive case, \(0<\beta<1\), at \(\bar t=1200\)
    (\(t/\tau_{\mathrm{PV}}\simeq0.43\) for the parameters of
    \S\ref{sec:Simulation}).
    (a) Interface geometry in the \((\bar x,\bar y)\) plane. The magenta curve
    is the numerically tracked interface obtained by advecting the initial
    line \(\bar x_0=0\) with the full effective velocity field. The cyan
    dashed curve is the Poiseuille envelope \(\bar Y_{\mathrm F}\), and the
    thin solid curves are representative constraint curves from
    \eqref{eq:repulsive_constraint}, labelled by their initial streamline
    \(\bar y_0\geq\bar y_{\mathrm c}\). The yellow curve is the branch
    generated by \(\bar y_0=\bar y_{\mathrm c}\), which defines the upper
    diffusiophoretic envelope \(\bar Y_{\mathrm{DP,R}}\). The black curve is
    the composite approximation \(\bar Y_{\mathrm R}\).
    (b) Analytical reconstruction \(n_{\mathrm R}\).
    (c) Pure-advection simulation of \eqref{eq:n_advection}.
    (d) Full advection--diffusion simulation of \eqref{eq:n_ade}.
    All density panels use the same colour scale, with saturation for
    \(n\geq2\).
    }
    \label{fig:Repulsive}
\end{figure*}

Figure~\ref{fig:Repulsive}(a) compares the numerically tracked interface with
the Poiseuille envelope \eqref{eq:YF} and the constraint curves
\eqref{eq:repulsive_constraint}. The tracked interface lies between the
Poiseuille and diffusiophoretic envelopes, and we approximate it by the
geometric closure
\begin{equation}
\bar Y_{\mathrm R}(\bar x,\bar t)=
\begin{cases}
\max\!\left[
\bar Y_{\mathrm F}(\bar x,\bar t),
\bar Y_{\mathrm{DP,R}}(\bar x,\bar t)
\right],
& -\bar t\leq \bar x\leq 0,\\[7pt]
\bar Y_{\mathrm F}(\bar x,\bar t)\,
\dfrac{\bar Y_{\mathrm{DP,R}}(\bar x,\bar t)}
      {\bar Y_{\mathrm{DP,R}}(\infty,\bar t)},
& 0<\bar x\leq \bar t/2 .
\end{cases}
\label{eq:YR}
\end{equation}
Behind the front centre \((\bar x\leq0)\), the interface is set by whichever
of the two envelopes lies higher. Ahead of the front centre
\((0<\bar x\leq\bar t/2)\), the multiplicative normalization in the second
branch ensures that the interface matches the Poiseuille envelope far ahead of
the front, where \(\bar Y_{\mathrm{DP,R}}(\infty,\bar t)=\bar y_{\mathrm c}\).
As in the attractive case, \eqref{eq:YR} is a geometric closure rather than an
exact characteristic solution.

The particle density inside the region reached by characteristics follows
from \eqref{eq:n_positiveX0}. Defining
\begin{equation}
g_{\mathrm R}(\bar x,\bar t)
=
n_0
\left[
\frac{\beta+1}{2}
-
\frac{\beta-1}{2}
\erf\left(\frac{\bar x}{\sqrt{4\bar t}}\right)
\right]^{-\bar\Gamma_{\mathrm p}},
\label{eq:g_R}
\end{equation}
the repulsive-case particle field is
\begin{equation}
n_{\mathrm R}(\bar x,\bar y,\bar t)=
\begin{cases}
g_{\mathrm R}(\bar x,\bar t),
& \bar x>\tfrac{\bar t}{2},\quad 0\leq\bar y\leq1,\\[3pt]
g_{\mathrm R}(\bar x,\bar t),
& -\bar t\leq\bar x\leq\tfrac{\bar t}{2},\quad
  \bar y\geq\bar Y_{\mathrm R}(\bar x,\bar t),\\[3pt]
0,
& \text{otherwise}.
\end{cases}
\label{eq:nField_xy_R}
\end{equation}

Figures~\ref{fig:Repulsive}(b--d) show that this reconstruction captures the
main large-scale features of the repulsive front: the upward displacement of
the interface and the formation of a near-wall particle-rich trailing region.
The full advection--diffusion simulation smooths the sharp non-diffusive
interface and broadens the near-wall plume but retains the same qualitative
structure.

\subsection{\label{subsec:Validation}Model validation}

We now compare the analytical reconstructions with the simulations of
\S\ref{sec:Simulation}. Comparison with the non-diffusive advection model
\eqref{eq:n_advection} tests the geometric closure used to reconstruct the
particle-occupied region; comparison with the full advection--diffusion
simulation \eqref{eq:n_ade} additionally tests the effect of particle diffusion
and the neglected streamwise diffusiophoretic drift on the predicted
distribution.

To compare in the laboratory frame, we transform from the moving frame using
\[
  \bar x=\bar x_{\mathrm{lab}}-\bar t,
  \qquad
  \bar x_{\mathrm{lab}}=\frac{x_{\mathrm{lab}}}{U_{\mathrm m}\tau_{\mathrm c}},
\]
where \(\bar x_{\mathrm{lab}}\) is measured from the inlet. The
cross-sectionally averaged particle density is
\begin{equation}
\langle n_i\rangle(\bar x_{\mathrm{lab}},\bar t)
=
\int_0^1
n_i(\bar x_{\mathrm{lab}}-\bar t,\bar y,\bar t)\,d\bar y,
\qquad
i\in\{\mathrm A,\mathrm R,\mathrm C\}.
\label{eq:n_average_def}
\end{equation}
For the attractive and repulsive cases, substituting
\eqref{eq:nField_xy_A} or \eqref{eq:nField_xy_R} gives
\begin{equation}
\langle n_i\rangle(\bar x_{\mathrm{lab}},\bar t)
=
\begin{cases}
0,
& \bar x_{\mathrm{lab}}<0,\\[3pt]
g_i(\bar x_{\mathrm{lab}}-\bar t,\bar t)
\left[
1-\bar Y_i(\bar x_{\mathrm{lab}}-\bar t,\bar t)
\right],
& 0\leq \bar x_{\mathrm{lab}}\leq \tfrac{3}{2}\bar t,\\[3pt]
g_i(\bar x_{\mathrm{lab}}-\bar t,\bar t),
& \bar x_{\mathrm{lab}}>\tfrac{3}{2}\bar t,
\end{cases}
\qquad
i\in\{\mathrm A,\mathrm R\},
\label{eq:nx_Lab}
\end{equation}
where the factor \(1-\bar Y_i\) is the fraction of the cross-section above the
particle interface. The interval
\(0\leq\bar x_{\mathrm{lab}}\leq3\bar t/2\) corresponds to the sheared particle
front, since the moving-frame front spans \(-\bar t\leq\bar x\leq\bar t/2\).
For the control case, the same expression applies with
\[
  g_{\mathrm C}=n_0,
  \qquad
  \bar Y_{\mathrm C}=\bar Y_{\mathrm F},
\]
because there is no diffusiophoretic migration.

Figure~\ref{fig:Simulation}(g--i) compares the analytical prediction
\eqref{eq:nx_Lab}, the non-diffusive advection model \eqref{eq:n_advection},
and the full advection--diffusion simulation \eqref{eq:n_ade}. The control case
provides the baseline response: in the absence of a solute contrast, the front
is broadened only by the parabolic Poiseuille profile, and particle diffusion
remains weak over the times shown because
\(\tau_{\mathrm p}\gg\tau_{\mathrm{PV}}\).

For the attractive case, the non-diffusive and full simulations agree closely,
confirming that particle diffusion and streamwise diffusiophoretic drift are
subdominant over the main displacement interval. The analytical model captures
the sharp density front and the accelerated particle removal. Its main
discrepancy is the leading peak ahead of the solute front, which corresponds to
the centreline high-density plume discussed in \S\ref{sec:Attractive}; this is
expected, because the plume is generated by particles initially close to the
solute interface, whereas the analytical reduction assumes
\(|\bar x_0|\gg1\).

For the repulsive case, the analytical prediction agrees well with the
non-diffusive advection model and captures the dominant density drop and
near-wall accumulation. The full advection--diffusion simulation shows a
smoother interface and a broader trailing structure: particle diffusion,
although slow compared with solute diffusion, becomes important near the sharp
interfaces and near-wall accumulations generated by the non-diffusive dynamics.

We also compare the total number of particles remaining in the channel. The
residual particle fraction is
\begin{equation}
\frac{N_i(\bar t)}{N_0}
=
\frac{1}{n_0\bar L}
\int_0^{\bar L}
\langle n_i\rangle(\bar x_{\mathrm{lab}},\bar t)\,d\bar x_{\mathrm{lab}},
\qquad
\bar L=\frac{L}{U_{\mathrm m}\tau_{\mathrm c}},
\qquad
N_0=n_0\bar L.
\label{eq:N_t}
\end{equation}
Figure~\ref{fig:Simulation}(f) compares \eqref{eq:N_t} with the non-diffusive
and full simulations. The analytical prediction is in close quantitative
agreement with the non-diffusive model for both attractive and repulsive cases,
indicating that the geometric closure captures the dominant transport
geometry. At later times, both non-diffusive descriptions overestimate the
residual tail relative to the full advection--diffusion simulations because
they do not include diffusive leakage across the sharp particle front.

The close agreement between the analytical and non-diffusive descriptions
confirms that the dominant transport physics in this regime is reachability:
which streamlines can be reached by particle characteristics, and how the local
density is amplified along them by the solute compression. Particle diffusion
modifies this picture primarily through smoothing of the sharp interfaces.
Across all three approaches, attractive diffusiophoresis produces substantially
faster particle removal over the first residence time, while repulsive
diffusiophoresis retains particles longer by pushing them toward the slow near-wall streamlines.

\section{Particle transport based on the large-deviation model}
\label{sec:ld-particle-transport}

The particle model in \S\ref{sec:Model} uses the Taylor-core solute field,
which is asymptotically valid for \(\tilde t\gg1\) in the distribution core,
\(\tilde x=O(\tilde t^{1/2})\), or equivalently
\(\zeta=\tilde x/\tilde t=O(\tilde t^{-1/2})\). Non-diffusive particles,
however, can travel over \(\tilde x=O(\tilde t)\) distances and may therefore
sample the finite-\(\zeta\) non-Gaussian tails, where the Taylor
approximation can fail. The validation in \S\ref{subsec:Validation} used
\(\mathrm{Pe}_{\mathrm s}=2\), for which these departures are weak. Here we
perform a more stringent test at \(\mathrm{Pe}_{\mathrm s}=2000\), where
they are much more pronounced \citep{haynes2014dispersion}.

We compare trajectories driven by the large-deviation field
\(\theta_{\mathrm{LD}}\) \citep{haynes2014dispersion}, evaluated using the finite-\(q\) reconstruction
\eqref{eq:ld_step_q_integral}, with trajectories driven by its Taylor-core
limit \(\theta_{\mathrm T}\), given by
\eqref{eq:ld_theta_T_particle}. For
\(m\in\{\mathrm{LD},\mathrm T\}\), define
\begin{equation}
  C_m(\tilde x,\bar y,\tilde t)
  =
  \frac{c_m(\tilde x,\bar y,\tilde t)}{c_{\mathrm i}}
  =
  1+(\beta-1)\theta_m(\tilde x,\bar y,\tilde t),
  \qquad
  \beta=\frac{c_{\mathrm f}}{c_{\mathrm i}}.
  \label{eq:ld_particle_Cm}
\end{equation}
The non-diffusive particle characteristics are
\begin{subequations}
\label{eq:ld_particle_characteristics}
\begin{align}
  \frac{\mathrm d\tilde x_m}{\mathrm d\tilde t}
  &=
  \mathrm{Pe}_{\mathrm s}v(\bar y_m)
  +
  \bar\Gamma_{\mathrm p}
  \frac{\partial}{\partial\tilde x}
  \ln C_m(\tilde x_m,\bar y_m,\tilde t),
  \label{eq:ld_particle_characteristic_x}
  \\
  \frac{\mathrm d\bar y_m}{\mathrm d\tilde t}
  &=
  \bar\Gamma_{\mathrm p}
  \frac{\partial}{\partial\bar y}
  \ln C_m(\tilde x_m,\bar y_m,\tilde t),
  \label{eq:ld_particle_characteristic_eta}
\end{align}
\end{subequations}
where
\[
  v(\bar y)=\frac12(1-3\bar y^2),
  \qquad
  \bar\Gamma_{\mathrm p}
  =
  \frac{\Gamma_{\mathrm p}}{D_{\mathrm s}}.
\]
Using the same characteristic equations for both solute fields ensures that
their difference isolates the non-Gaussian correction. Both integrations
retain the streamwise diffusiophoretic velocity and the temporal broadening
of the solute front.

Trajectories are initialized at a late time \(\tilde t_0\gg1\):
\begin{equation}
  \tilde x_m(\tilde t_0)=\tilde x_0>0,
  \qquad
  \bar y_m(\tilde t_0)=\bar y_0,
  \qquad
  \bar y_{\mathrm c}<\bar y_0\leq1,
  \qquad
  \bar y_{\mathrm c}=\frac{1}{\sqrt3}.
  \label{eq:ld_particle_initial_conditions}
\end{equation}
Since \(v(\bar y_0)<0\) for \(\bar y_0>\bar y_{\mathrm c}\), these particles
move toward decreasing \(\tilde x\) and encounter the front from downstream.
We choose \(\tilde x_0\) sufficiently large that
\[
  \frac{U_{\mathrm m}a}{D_{\mathrm{s,eff}}}\tilde x_0\gg1,
  \qquad
  \frac{\tilde x_0}
       {\sqrt{\bar D_{\mathrm{s,eff}}\tilde t_0}}
  \gg1.
\]
The first condition is the far-field condition underlying
\eqref{eq:invariant_y}; the second places the particle several front widths
downstream, where its initial diffusiophoretic velocity is negligible.

After interacting with the front, a particle either enters the invading
fluid or, in the attractive case, may cross the mean-flow streamline and
returns to the original fluid. Each trajectory is continued until the
diffusiophoretic velocity becomes negligible. Its final transverse position
is defined by
\begin{equation}
  \bar y_{1,m}(\bar y_0)
  =
  \lim_{\tilde t\to\infty}\bar y_m(\tilde t),
  \qquad
  m\in\{\mathrm{LD},\mathrm T\}.
  \label{eq:ld_eta1_numerical}
\end{equation}
A transmitted trajectory approaches \(C_m=\beta\) and
\(\tilde x\to-\infty\), whereas a reflected trajectory returns to
\(C_m=1\) and \(\tilde x\to+\infty\).

For comparison, we also use the far-field endpoint map implied by the
approximate invariant \eqref{eq:invariant_y}. As in
\S\ref{sec:Invariants}, this map neglects streamwise diffusiophoresis and
the front-broadening remainder \(\mathcal I\). Let
\(F_0=F(\bar y_0)\), with \(F\) defined in \eqref{eq:F_def}. Since
\(C_\beta=1\) in the initial far field, while
\(C_\beta\to\beta\) for a transmitted trajectory and
\(C_\beta\to1\) for a reflected trajectory, the analytical endpoint obeys
\[
  F(\bar y_{1,\mathrm{an}})
  =
  \begin{cases}
    \beta^{\bar\Gamma_{\mathrm p}}F_0,
    & \text{transmitted},\\
    F_0,
    & \text{reflected}.
  \end{cases}
\]

For \(0\leq z\leq1\), let \(F_{\mathrm w}^{-1}\) and
\(F_{\mathrm c}^{-1}\) denote the wall-side and centreline-side inverse
branches of \(F\), respectively:
\begin{equation}
  \begin{aligned}
  F_{\mathrm w}^{-1}(z)
  &=
  \frac{2}{\sqrt3}
  \cos\!\left[
    \frac{\pi-\cos^{-1}z}{3}
  \right],
  \\
  F_{\mathrm c}^{-1}(z)
  &=
  -\frac{2}{\sqrt3}
  \cos\!\left[
    \frac{2\pi-\cos^{-1}z}{3}
  \right].
  \end{aligned}
  \label{eq:ld_F_inverse_branches}
\end{equation}
Here
\(F_{\mathrm w}^{-1}(z)\in[\bar y_{\mathrm c},1]\) and
\(F_{\mathrm c}^{-1}(z)\in[0,\bar y_{\mathrm c}]\).

For \(0<\beta<1\), all trajectories considered here are transmitted and
remain on the wall-side branch. Their analytical endpoints are therefore
\begin{equation}
  \bar y_{1,\mathrm{an}}(\bar y_0)
  =
  F_{\mathrm w}^{-1}
  \!\left(
    \beta^{\bar\Gamma_{\mathrm p}}F_0
  \right),
  \qquad
  \bar y_{\mathrm c}<\bar y_0\leq1,
  \quad
  0<\beta<1.
  \label{eq:ld_eta1_repulsive_analytical}
\end{equation}

For \(\beta>1\), the analytical reflected--transmitted threshold is the
wall-side separatrix \(\bar y_{\mathrm a1}\) introduced in
\eqref{eq:ya_roots}, or equivalently, 
\(
  \bar y_{\mathrm a1}
  =
  F_{\mathrm w}^{-1}
  \!\left(\beta^{-\bar\Gamma_{\mathrm p}}\right).
\)
The analytical endpoint map is
\begin{equation}
  \bar y_{1,\mathrm{an}}(\bar y_0)
  =
  \begin{cases}
  F_{\mathrm c}^{-1}(F_0),
  &
  \bar y_{\mathrm c}<\bar y_0<\bar y_{\mathrm a1}
  \quad\text{(reflected)},
  \\[5pt]
  F_{\mathrm w}^{-1}
  \!\left(
    \beta^{\bar\Gamma_{\mathrm p}}F_0
  \right),
  &
  \bar y_{\mathrm a1}<\bar y_0\leq1
  \quad\text{(transmitted)}.
  \end{cases}
  \qquad
  \beta>1.
  \label{eq:ld_eta1_attractive_analytical}
\end{equation}
The limiting separatrix trajectory with
\(\bar y_0=\bar y_{\mathrm a1}\) approaches
\(\bar y_{\mathrm c}\) asymptotically and has no regular far-field
endpoint.

We compare
\(\bar y_{1,\mathrm{LD}}(\bar y_0)\),
\(\bar y_{1,\mathrm T}(\bar y_0)\), and
\(\bar y_{1,\mathrm{an}}(\bar y_0)\).
To identify where the transverse migration occurs, define the fraction of
the net displacement acquired inside a front-centred window:
\begin{equation}
  \mathcal F_m(k;\bar y_0)
  =
  \frac{
    \left|
    \displaystyle
    \int_{\tilde t_0}^{\infty}
    H\!\left(
      -|\tilde x_m|
      +
      k\sqrt{2\bar D_{\mathrm{s,eff}}\tilde t}
    \right)
    \frac{\mathrm d\bar y_m}{\mathrm d\tilde t}
    \,\mathrm d\tilde t
    \right|
  }{
    \left|
      \bar y_{1,m}-\bar y_0
    \right|
  },
  \qquad
  m\in\{\mathrm{LD},\mathrm T\}.
  \label{eq:front_window_fraction}
\end{equation}
Here, \(H\) denotes the Heaviside step function. In the numerical quadrature, each trajectory
segment is assigned to the window using its midpoint. Thus
\(\mathcal F_m(k;\bar y_0)\simeq1\) means that nearly all of the net
transverse migration occurs within the specified window.

\begin{figure*}
  \centering
  \includegraphics[width=\textwidth,keepaspectratio]{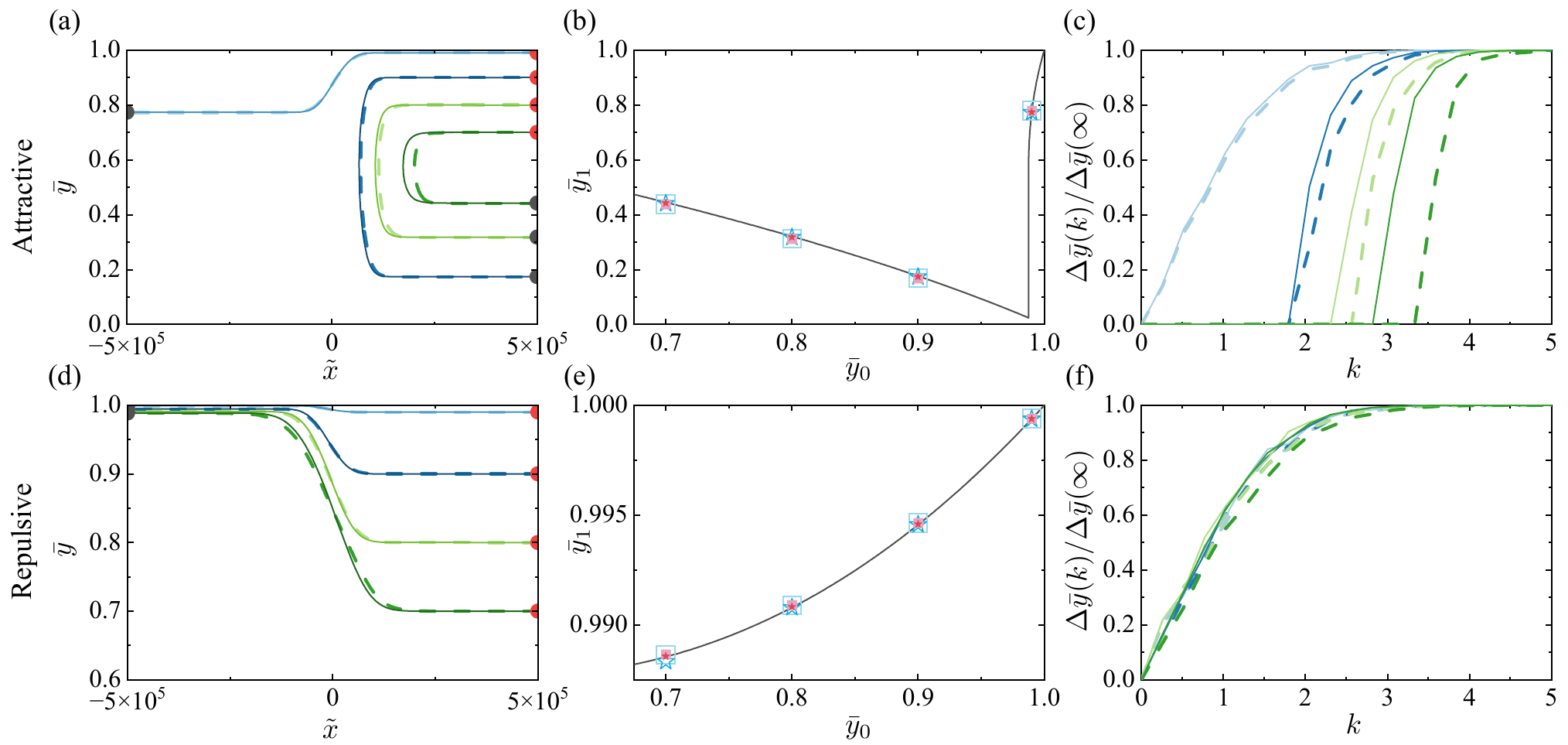}
  \captionsetup{
    width=\textwidth,
    justification=justified,
    singlelinecheck=false
  }
  \caption{
  Particle scattering by attractive
  (\(\beta=100\), panels a--c) and repulsive
  (\(\beta=0.01\), panels d--f) solute fronts.
  (a,d) Representative trajectories at
  \(\mathrm{Pe}_{\mathrm s}=2000\) for
  \(\bar y_0\in\{0.7,0.8,0.9,0.99\}\).
  Solid and dashed curves use the Taylor-core field and the finite-\(q\)
  large-deviation reconstruction, respectively; red and black markers
  denote initial and final positions.
  (b,e) Numerical endpoint maps computed from the large-deviation field
  (open symbols) and Taylor-core field (filled symbols) at
  \(\mathrm{Pe}_{\mathrm s}=2000\) (stars) and
  \(\mathrm{Pe}_{\mathrm s}=2\) (squares). Black curves show the analytical
  maps \eqref{eq:ld_eta1_attractive_analytical} and
  \eqref{eq:ld_eta1_repulsive_analytical}.
  (c,f) Fraction \(\mathcal F_m(k;\bar y_0)\) of the transverse displacement
  acquired within
  \(|\tilde x|<k\sqrt{2\bar D_{\mathrm{s,eff}}\tilde t}\) at
  \(\mathrm{Pe}_{\mathrm s}=2000\).
  }
  \label{fig:y1y0_Mapping}
\end{figure*}

Figure~\ref{fig:y1y0_Mapping}(a,d) shows that the large-deviation and
Taylor-core trajectories remain close even at
\(\mathrm{Pe}_{\mathrm s}=2000\). They differ slightly while crossing the
region of strongest solute gradient but approach nearly identical final
transverse positions.

The endpoint maps in figure~\ref{fig:y1y0_Mapping}(b,e) confirm this
result. At both values of \(\mathrm{Pe}_{\mathrm s}\) and for both
concentration contrasts, the large-deviation and Taylor-core endpoints
closely follow the analytical invariant maps. The non-Gaussian tails
therefore have little effect on the net cross-streamline displacement.

Panels~\ref{fig:y1y0_Mapping}(c,f) explain this insensitivity. Nearly all
transverse migration occurs within a few front widths,
\[
  |\tilde x|
  <
  k\sqrt{2\bar D_{\mathrm{s,eff}}\tilde t},
  \qquad
  k=O(1).
\]
This window has Taylor-core scaling:
\[
  \tilde x=O(\tilde t^{1/2}),
  \qquad
  \zeta=O(\tilde t^{-1/2}),
  \qquad
  q=O(\tilde t^{-1/2}).
\]
The large-deviation field therefore reduces locally to its Taylor-core
approximation where most of the migration occurs. Outside this region, the
concentration gradients are too weak to contribute appreciably.

Thus, even at \(\mathrm{Pe}_{\mathrm s}=2000\), finite-\(\zeta\)
corrections do not significantly alter the final transverse displacement.
This supports the use of the Taylor-core solute field and the approximate
invariant map to predict net cross-streamline migration in the
non-diffusive particle-transport regime considered in \S \ref{sec:Model}.

\section{\label{sec:Discussion}Discussion}

The central implication of this work is that a one-dimensional
Taylor-dispersion description of the solute, while accurate for solute
transport in pressure-driven channel flow, can be qualitatively misleading for
the colloid transport that the solute drives. Diffusiophoresis depends on
\(\nabla\ln c\), not on \(c\) itself, and the residual transverse correction to
the Taylor-dispersed solute field---small in the concentration but
P\'eclet-enhanced in the gradient---is sufficient to redistribute particles
between streamlines and thereby reorganize their longitudinal transport. For the two-dimensional
plane-Poiseuille channel considered here, the resulting leading-order
transverse diffusiophoretic drift is
\begin{equation}
    V_y
    \approx
    \frac{1}{2}\Gamma_{\mathrm p}\,\mathrm{Pe}_{\mathrm s}\,
    \frac{y}{a}
    \left[1-\left(\frac{y}{a}\right)^2\right]
    \partial_x\ln C_\beta.
\end{equation}
This drift vanishes at the centreline and the wall but remains finite in the
channel interior, where it redistributes particles between streamlines and
thereby reorganizes their longitudinal transport.

This is important as a broad separation between solute and particle transverse
diffusion times is common for colloids in confined flows. Molecular solutes
equilibrate across micron-scale gaps on sub-second timescales, whereas
micron-scale colloids can remain weakly diffusive across the gap over the
residence times relevant to breakthrough and removal measurements. Many
microfluidic, filtration, and porous-media experiments therefore operate in
the intermediate window identified here, rather than in either the early
fully two-dimensional limit or the ultimate macrotransport limit. In this
window, the solute may be nearly one-dimensional in concentration but remains
two-dimensional in gradient, and any description that retains only the former
will miss the cross-streamline particle response generated by the latter.

This mechanism offers a natural interpretation of the strong
diffusiophoretic signatures recently observed in preferential flow pathways of
porous media \citep{alipour2026diffusiophoretic, Pujari26, Pahlavan26}. Dimensional estimates based
on transverse solute equilibration suggest that solute gradients should be
confined either to dead-end pores or to the streamwise direction in conducting
channels, and that diffusiophoresis should therefore be weak in the flow
pathways themselves. The cross-streamline migration mechanism identified here
removes this apparent paradox: even a nominally one-dimensional
Taylor-dispersed solute front carries a residual transverse gradient in a
sheared flow, and this gradient can move colloids between slow and fast
streamlines.

More broadly, the result adds to a growing body of work showing that weak
phoretic drifts can be amplified by flow structure to produce macroscopic
transport consequences. In chaotic and cellular flows, salt-driven phoretic
motion can enhance or suppress colloid mixing, generate effective
compressibility, and even block transport between flow cells
\citep{deseigne2014pinch,volk2014chaotic,raynal2018advection,volk2022phoresis}.
In porous and complex media, pore-scale disorder, local correlations, and
fluid deformation control the transition from microscopic flow structure to
macroscopic dispersion and mixing
\citep{meigel2022dispersive,dentz2023mixing,leborgne2026fluid}. In biological
porous media, chemical gradients coupled to flow disorder can strongly affect
the spatial distribution and residence time of motile cells
\citep{deanna2021chemotaxis,scheidweiler2024spatial}. The present results add
a complementary passive-colloid mechanism: chemical gradients can modulate the
effect of shear and geometric heterogeneity not by changing the flow field, but
by moving particles across streamlines. In disordered media, this coupling may
either enhance or suppress the influence of geometric heterogeneity, depending
on the sign of the solute contrast, the geometry of the conducting backbone,
and the particle mobility. Extending the present mechanism to pore networks
and fully resolved disordered geometries is therefore a natural next step.

The cross-streamline migration identified here is distinct from other known
routes by which particles move laterally in channel flows. Inertial migration
and inertial microfluidic focusing arise from finite-Reynolds-number lift
forces and lead to equilibrium focusing positions set by the balance of shear-
gradient and wall-induced lift forces
\citep{segre1961radial,ho1974inertial,dicarlo2009inertial,
amini2014inertial}. Shear-induced migration in concentrated suspensions,
by contrast, results from particle--particle interactions and normal-stress or
collision-induced fluxes that drive particles away from regions of large shear
rate or large particle concentration
\citep{leighton1987migration,nott1994pressure,phillips1992constitutive,
morris1999curvilinear}. Lateral displacement can also be imposed by
external or engineered fields, including acoustic, electric, magnetic, or
geometric forcing in microfluidic separation devices
\citep{lenshof2010continuous,pethig2010dielectrophoresis,
pamme2006magnetism,huang2004continuous}. The mechanism studied here does
not rely on inertia, particle collisions, finite concentration, imposed
external fields, or structured obstacles. Instead, it arises in dilute
low-Reynolds-number channel flow from a residual transverse chemical gradient
that persists in a Taylor-dispersed solute field and drives diffusiophoretic
migration across streamlines.

Although all experiments and simulations presented here employ a positive
diffusiophoretic mobility, \(\Gamma_{\mathrm p}>0\), the cross-streamline
migration mechanism and the governing equations developed in
\S\ref{sec:Model} do not rely on this restriction. With the convention
\(\mathbf{u}_{\mathrm{DP}}=\Gamma_{\mathrm p}\nabla\ln c\) used in
\eqref{eq:uDP}, a negative mobility corresponds to particle migration toward
lower solute concentration. For a dilute binary electrolyte, the classical
thin-double-layer theory decomposes the diffusiophoretic mobility into
chemiphoretic and electrophoretic contributions
\citep{prieve_motion_1984}. The chemiphoretic contribution arises from the osmotic-pressure imbalance associated with the asymmetric distribution of ions within the electrical
double layer and, within this classical limit, drives particles toward higher
electrolyte concentration. The electrophoretic contribution is generated by
the diffusion electric field arising from unequal cation and anion
diffusivities, and its sign depends on the particle \(\zeta\)-potential and
the ionic diffusivity contrast. When the electrophoretic contribution opposes
and exceeds the chemiphoretic contribution, the net diffusiophoretic mobility
becomes negative. Additional mobility reversals can arise at finite
double-layer thickness and sufficiently large \(\lvert\zeta\rvert\)
\citep{prieve1987diffusiophoresis}, and negative mobilities have been observed
experimentally for polystyrene particles in carboxylic-acid gradients
\citep{timmerhuis2022diffusiophoretic}. The present governing framework
therefore extends directly to \(\Gamma_{\mathrm p}<0\); for a fixed
concentration contrast, the direction of cross-streamline migration is
reversed, interchanging the attractive and repulsive responses described
above.

The most direct extension of the present theory is to incorporate particle
diffusion systematically. In the experiments and full advection--diffusion
simulations, diffusion smooths the sharp interfaces predicted by the
non-diffusive model and broadens the near-wall accumulation in the repulsive
case. Over longer times, particle diffusion must also drive the system toward
the macrotransport limit. A theory that retains weak particle diffusion would
therefore interpolate between the non-diffusive description developed here and
the late-time macrotransport regime. A second direction is to relax the
no-slip/no-flux idealization at the walls. Diffusio-osmotic slip driven by the
same solute gradients can modify near-wall transport and produce Taylor-like
dispersion of solutes and particles \citep{alessio2022diffusioosmosis,teng2023diffusioosmotic,Chakra23,Migacz23,Liu25}. Such wall-driven flows could reinforce or oppose the
diffusiophoretic cross-streamline migration, depending on surface chemistry.

Another extension is to incorporate concentration-dependent
diffusiophoretic mobility into the analytical framework. The numerical tests
in \S\ref{sec:cEffect}, which include finite-double-layer effects and a
concentration-dependent particle \(\zeta\)-potential, show that
\(\Gamma_{\mathrm p}(c)\) modifies the degree of particle-front sharpening and
broadening without changing the qualitative transport patterns over the
parameter range examined. A theory that retains \(\Gamma_{\mathrm p}(c)\)
would be particularly valuable in regimes where the mobility varies strongly
with concentration or changes sign.

A further extension is to examine diffusiophoretic transport in
channel flows containing multicomponent electrolyte solutions
\citep{khair2023dispersion}. In such solutions, unequal ionic diffusivities
generate diffusion potentials that couple the transport of the different
species. The work of \cite{ding2023shear} showed that, for more than two ionic species,
the long-time channel-scale dynamics are generally governed by a nonlinear
effective transport equation with a concentration-dependent diffusion tensor,
rather than by independent advection--diffusion equations. This coupling can
produce non-Gaussian concentration profiles, spontaneous separation of ionic
species, and even upstream migration. Determining how these coupled ionic
concentration gradients influence colloidal diffusiophoresis is therefore a
natural direction for future study
\citep{gupta2019diffusiophoretic}. The wall-induced corrections to particle diffusivity could also be incorporated within the same framework \citep{Gupta20,shim2022diffusiophoresis,Lee23,ding2023shear}.



\backsection[Funding]{We acknowledge Office of the Under Secretary of Defense for Research and Engineering (FA9550-22-1-0320), and National Science Foundation CAREER award (2443484) for partial support of this work. }

\backsection[Declaration of interests]{The authors report no conflict of interest.}




\appendix
\setcounter{figure}{0}
\renewcommand{\thefigure}{A\arabic{figure}}

\appendix

\section{Estimate of the remainder term}\label{app:integral}

\subsection{Asymptotic estimate}

In this appendix we estimate the remainder term appearing in
\eqref{eq:invariant_y_full} and \eqref{eq:invariant_phi_full},
\begin{equation}
R(\bar t;\bar x_0,\bar y_0)
:=\bar\Gamma_{\mathrm p}\int_0^{\bar t}
\partial_{\bar t}\ln \bar S(\bar x(s),s;\beta)\,\mathrm ds,
\label{eq:Rdef}
\end{equation}
where $(\bar x(s),\bar y(s))$ is a characteristic satisfying
\eqref{eq:char_x}--\eqref{eq:char_y} with initial data
$(\bar x_0,\bar y_0)$ at $s=0$.
Our goal is to show that
\begin{equation}
R(\bar t;\bar x_0,\bar y_0)=\mathcal{O}(|\bar x_0|^{-1/2})
\qquad\text{for }|\bar x_0|\gg1,
\label{eq:Rgoal}
\end{equation}
uniformly over the times relevant to the reduced model.

We begin from the explicit form
\begin{equation}
\bar X(\bar x,\bar t)=\frac12\erfc\!\left(\frac{\bar x}{2\sqrt{\bar t}}\right),
\qquad
\bar S(\bar x,\bar t;\beta)=
\begin{cases}
\bar X+(\beta-1)^{-1}, & \beta>1,\\[2pt]
-\bar X+(1-\beta)^{-1}, & 0<\beta<1.
\end{cases}
\label{eq:XS_appendix}
\end{equation}
Differentiating $\bar X$ at fixed $\bar x$ gives
\begin{equation}
\partial_{\bar t}\bar X(\bar x,\bar t)
=
\frac{\bar x}{4\sqrt{\pi}\,\bar t^{3/2}}
\exp\!\left(-\frac{\bar x^2}{4\bar t}\right).
\label{eq:dtdX}
\end{equation}
Hence
\begin{equation}
\partial_{\bar t}\ln \bar S
=
\frac{\sigma_\beta\,\partial_{\bar t}\bar X}{\bar S},
\qquad
\sigma_\beta=
\begin{cases}
1, & \beta>1,\\
-1, & 0<\beta<1.
\end{cases}
\label{eq:dtdlogS1}
\end{equation}
Moreover, for fixed $\beta\neq1$, $\bar S$ is bounded away from zero:
\begin{equation}
\bar S(\bar x,\bar t;\beta)\ge
\begin{cases}
(\beta-1)^{-1}, & \beta>1,\\[2pt]
\beta(1-\beta)^{-1}, & 0<\beta<1.
\end{cases}
\label{eq:Spositive}
\end{equation}
Therefore there exists a constant $K_\beta>0$, depending only on $\beta$, such that
\begin{equation}
\bigl|\partial_{\bar t}\ln \bar S(\bar x,\bar t;\beta)\bigr|
\le
K_\beta\,
\frac{|\bar x|}{\bar t^{3/2}}
\exp\!\left(-\frac{\bar x^2}{4\bar t}\right).
\label{eq:dtdlogSbound}
\end{equation}
Substituting \eqref{eq:dtdlogSbound} into \eqref{eq:Rdef} yields
\begin{equation}
|R(\bar t;\bar x_0,\bar y_0)|
\le
K_\beta
\int_0^{\bar t}
\frac{|\bar x(s)|}{s^{3/2}}
\exp\!\left(-\frac{\bar x(s)^2}{4s}\right)\,\mathrm ds.
\label{eq:Rbasicbound}
\end{equation}

The kernel in \eqref{eq:Rbasicbound} is exponentially localized to the
diffusive front region
\begin{equation}
|\bar x|=\mathcal{O}(\sqrt{\bar t}).
\label{eq:frontlayer}
\end{equation}
Thus the remainder can almost only accumulate while the trajectory is inside this
$\mathcal{O}(\sqrt{\bar t})$-wide layer.

For characteristics not close to the
critical streamline
\begin{equation}
\bar y_{\mathrm c}=\frac{1}{\sqrt3},
\label{eq:yc_app}
\end{equation}
the streamwise speed
\begin{equation}
\dot{\bar x}=\bar V_x(\bar y)=\frac12(1-3\bar y^2)
\label{eq:Vx_app}
\end{equation}
is $\mathcal{O}(1)$. Since the layer width is $\mathcal{O}(\sqrt{\bar t})$, the residence time
inside the front layer is then $\mathcal{O}(\sqrt{\bar t})$. For particles released far
from the front, the front is encountered at times $s=\mathcal{O}(|\bar x_0|)$, so the
time spent in the front region is $\mathcal{O}(|\bar x_0|^{1/2})$.

The only possible complication is a trajectory passing very near
$\bar y=\bar y_{\mathrm c}$, where $\bar V_x=0$. However, the transverse
velocity does not vanish there. Indeed, from \eqref{eq:Ueff_xy},
\begin{equation}
\dot{\bar y}\big|_{\bar y=\bar y_{\mathrm c}}
=
\frac{\bar\Gamma_{\mathrm p}}{3\sqrt3}\,
\partial_{\bar x}\ln \bar S(\bar x,\bar t;\beta).
\label{eq:Vy_yc}
\end{equation}
Since $\partial_{\bar x}\bar X<0$, we have
\begin{equation}
\partial_{\bar x}\ln \bar S<0 \quad (\beta>1),
\qquad
\partial_{\bar x}\ln \bar S>0 \quad (0<\beta<1).
\label{eq:sign_dxdlogS}
\end{equation}
Thus, at $\bar y=\bar y_{\mathrm c}$,
\begin{equation}
\dot{\bar y}<0 \quad (\beta>1),
\qquad
\dot{\bar y}>0 \quad (0<\beta<1),
\label{eq:sign_Vy_yc}
\end{equation}
so the trajectory is pushed away from the critical streamline rather than
remaining trapped on it. In the front layer,
$\partial_{\bar x}\ln \bar S=\mathcal{O}(\bar t^{-1/2})$, and therefore
\begin{equation}
\dot{\bar y}\big|_{\bar y=\bar y_{\mathrm c}}=\mathcal{O}(\bar t^{-1/2}).
\label{eq:Vy_yc_scale}
\end{equation}
Over the time $\mathcal{O}(\sqrt{\bar t})$, this produces an
$\mathcal{O}(1)$ change in $\bar y$. Consequently, even trajectories that enter the
front layer arbitrarily close to $\bar y_{\mathrm c}$ are rapidly displaced to
regions where $\bar V_x=\mathcal{O}(1)$, so their residence time in $\mathcal F_M$
remains $\mathcal{O}(\sqrt{\bar t})$.

Inside the front layer, we have $|\bar x|=\mathcal{O}(\sqrt{\bar t})$, and therefore
\eqref{eq:dtdlogSbound} reduces to
\begin{equation}
\bigl|\partial_{\bar t}\ln \bar S\bigr|=\mathcal{O}(\bar t^{-1}).
\label{eq:frontamp}
\end{equation}
Since the front is encountered at $s=\mathcal{O}(|\bar x_0|)$, this gives
\begin{equation}
\bigl|\partial_{\bar t}\ln \bar S(\bar x(s),s;\beta)\bigr|
=\mathcal{O}(|\bar x_0|^{-1}).
\label{eq:frontamp_x0}
\end{equation}
Multiplying by the residence time $\mathcal{O}(|\bar x_0|^{1/2})$, we obtain
\begin{equation}
R(\bar t;\bar x_0,\bar y_0)=\mathcal{O}(|\bar x_0|^{-1/2}),
\qquad |\bar x_0|\gg1.
\label{eq:Rfinal}
\end{equation}

Equation \eqref{eq:Rfinal} justifies neglecting the remainder in
\eqref{eq:invariant_y_full} and \eqref{eq:invariant_phi_full} for particles
released sufficiently far from the solute front. The estimate is asymptotic, but it shows the relevant point for the reduced model:
the non-invariant part is generated only during the finite time that the
trajectory spends crossing the diffusive front, and that contribution decays
as $|\bar x_0|^{-1/2}$ for large $|\bar x_0|$.

\subsection{Numerical verification}

\begin{figure*}
    \centering
\includegraphics[width=\textwidth,keepaspectratio]{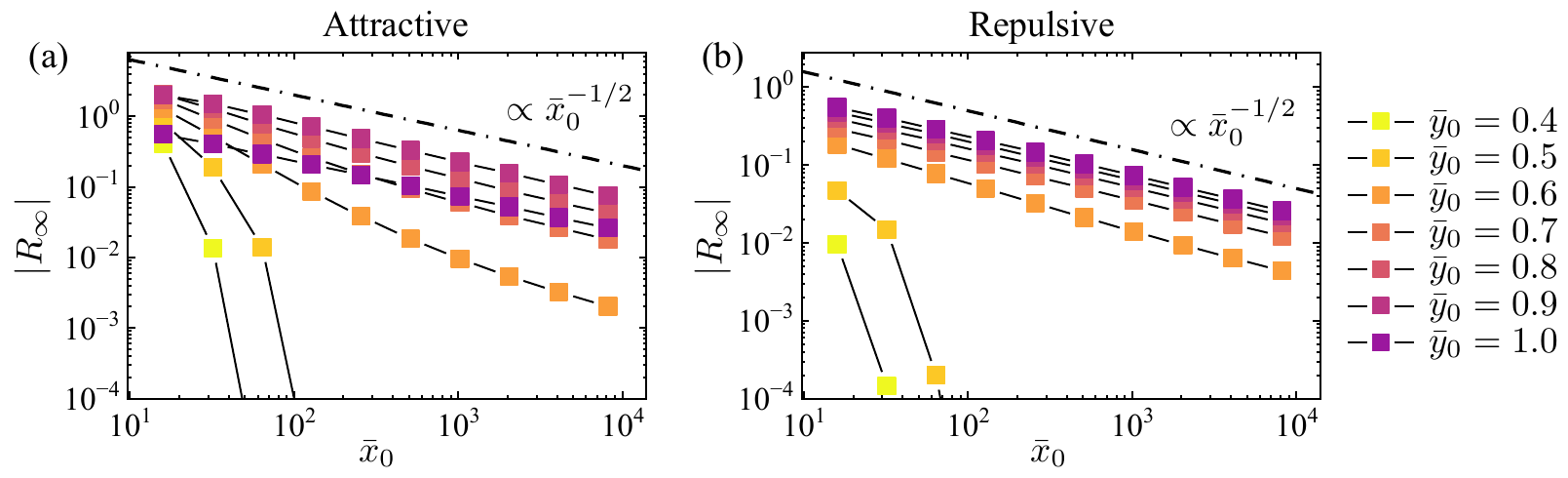}
\captionsetup{width=\textwidth,justification=justified,singlelinecheck=false}
    \caption{Numerical verification of the residual approximation. Panels
    (a,b) show the asymptotic residual magnitude \(|R_\infty|\) as a
    function of \(\bar x_0\) for the set of initial positions \(\bar y_0\)
    used in the parameter sweep, for the attractive (a) and repulsive (b)
    cases. The dash--dot line indicates the \(-1/2\) slope predicted by
    \eqref{eq:Rfinal}.}
    \label{fig:ResidualTermCheck}
\end{figure*}

In addition to the asymptotic estimate above, we numerically examine the
approximation underlying \eqref{eq:invariant_y} and
\eqref{eq:invariant_n}. To do so, we integrate the full
characteristic system \eqref{eq:char_x}--\eqref{eq:char_phi} together
with the residual evolution equation
\begin{equation}
\dot R
=\bar\Gamma_{\mathrm p}\,\partial_{\bar t}\ln \bar S(\bar x(\bar t),\bar t;\beta).
\end{equation}

We consider representative attractive and repulsive cases with
\(\beta=100\) and \(\beta=10^{-2}\), respectively, and fix
\(\bar\Gamma_{\mathrm p}=0.6\). The initial conditions are sampled over
\(\bar y_0=0.10,\,0.20,\,\ldots,\,1.00\)
and
\(\bar x_0=16,\,32,\,64,\,\ldots,\,8192\).
The integrations are started at \(\bar t_{\mathrm{start}}=10^{-3}\).
For each trajectory, we define the asymptotic residual
\begin{equation}
R_\infty:=R(\bar t_\infty;\bar x_0,\bar y_0),
\qquad
\bar t_\infty=16\,\bar t_{\mathrm{ref}},
\qquad
\bar t_{\mathrm{ref}}=\bar x_0/|\bar V_x(\bar y_0)|.
\end{equation}
The value of \(R\) at \(12\,\bar t_{\mathrm{ref}}\) is used as a
saturation check.

The resulting values of \(|R_\infty|\) for the attractive and repulsive
cases are shown in figure~\ref{fig:ResidualTermCheck}(a,b),
respectively. For trajectories on the front-interacting branch,
\(\bar y_0>\bar y_{\mathrm c}\), the numerical slopes are broadly
consistent with the \(|\bar x_0|^{-1/2}\) scaling predicted by
\eqref{eq:Rfinal}. By contrast, for trajectories released below the
critical streamline, \(\bar y_0<\bar y_{\mathrm c}\), the residual
decays much more rapidly, approximately exponentially with \(\bar x_0\).
These computations therefore support the asymptotic estimate
\eqref{eq:Rfinal} on the \(\bar y_0>\bar y_{\mathrm c}\) branch and,
more generally, justify neglecting the residual term in the approximate
invariants for particles released sufficiently far downstream of the
solute front.

\section{Effect of concentration-dependent diffusiophoretic mobility}
\label{sec:cEffect}

The main analysis assumes a constant diffusiophoretic mobility
\(\Gamma_{\mathrm p}\). In practice, the local salt concentration affects both
the particle zeta potential and the Debye length, and therefore the mobility.
To test the robustness of our conclusions, we evaluate
\(\Gamma_{\mathrm p}(c)\) using the finite-double-layer formulation of
\citet{prieve_motion_1984}, together with the fitted corrections and
concentration-dependent parameterization described by
\citet{lee_role_2023}. We used the same formulation in our previous study
\citep{Li_Alipour_Pahlavan_2026}.

We write the dimensionless local mobility as
\begin{equation}
\bar{\Gamma}_{\mathrm p}(c)
\equiv
\frac{\Gamma_{\mathrm p}(c)}{D_{\mathrm s}}
=
\frac{1}{D_{\mathrm s}}
\Gamma_{\mathrm p}
\!\left[\zeta(c),\lambda(c)\right],
\qquad
\lambda(c)=\frac{1}{\kappa(c)a_{\mathrm p}},
\label{eq:Gp(c)}
\end{equation}
where \(\Gamma_{\mathrm p}(\zeta,\lambda)\) denotes the
finite-double-layer mobility, \(a_{\mathrm p}\) is the particle radius, and
\(\kappa^{-1}\) is the Debye length. The complete finite-double-layer
expression and its fitted correction functions are given in
\citet{prieve_motion_1984,lee_role_2023}; implementation details for the same
formulation are also summarized in our previous application
\citep{Li_Alipour_Pahlavan_2026}.

Under the constant-surface-charge approximation, the concentration dependence
enters through
\begin{subequations}
\label{eq:Gp_concentration_relations}
\begin{align}
\zeta(c)
&=
a_0+a_1\log_{10}\!\left(\frac{Z^2c}{c_0}\right),
\label{eq:zeta_vs_C}
\\
\kappa^{-1}(c)
&=
\left(
\frac{\varepsilon k_B T}
{2Z^2e^2N_Ac_{\mathrm{SI}}}
\right)^{1/2},
\label{eq:Debye_c}
\end{align}
\end{subequations}
where \(Z\) is the electrolyte valence, \(\varepsilon\) is the permittivity of
the medium, \(k_B\) is Boltzmann's constant, \(T\) is the absolute
temperature, \(e\) is the elementary charge, and \(N_A\) is Avogadro's
constant. In \eqref{eq:zeta_vs_C}, \(c_0=1~\mathrm{M}\), with \(c\) and
\(c_0\) expressed in the same concentration units. In
\eqref{eq:Debye_c}, \(c_{\mathrm{SI}}\) is expressed in
\(\mathrm{mol\,m^{-3}}\); thus, when \(c\) is expressed in
\(\mathrm{mol\,l^{-1}}\), \(c_{\mathrm{SI}}=10^3c\).

For the representative latex--NaCl parameterization considered here, we use
\begin{equation}
a_0\simeq0~\mathrm{mV},
\qquad
a_1\simeq-43.7~\mathrm{mV\,decade^{-1}},
\label{eq:zeta_parameters}
\end{equation}
following
\citet{kirby_zeta_2004,kirby_zeta_2004-1},
\citet{staffeld_diffusion-induced_1989}, and
\citet{lee_role_2023}. Using this local mobility, the diffusiophoretic velocity in
\eqref{eq:uDP} is evaluated pointwise as
\begin{equation}
\mathbf{u}_{\mathrm{DP}}(x,y,t)
=
\Gamma_{\mathrm p}[c(x,y,t)]
\nabla\ln c(x,y,t).
\label{eq:uDP_variable}
\end{equation}

We repeat the simulations of \S\ref{sec:Simulation} for two concentration
pairs that span the same two-decade concentration contrast but differ in their
absolute salt levels. The higher-concentration pair is
\(0.1~\mathrm{mM}\) and \(10~\mathrm{mM}\), whereas the
lower-concentration pair is \(0.001~\mathrm{mM}\) and
\(0.1~\mathrm{mM}\). For the attractive displacement, the smaller
concentration in each pair is \(c_{\mathrm i}\) and the larger concentration
is \(c_{\mathrm f}\); these assignments are reversed for the repulsive
displacement. Thus,
\(c_{\mathrm f}/c_{\mathrm i}=100\) in both attractive cases and
\(c_{\mathrm f}/c_{\mathrm i}=10^{-2}\) in both repulsive cases. Since the
concentration ratio is preserved, the comparison isolates the influence of
the absolute salt concentration through \(\Gamma_{\mathrm p}(c)\).

\begin{figure*}
    \centering
    \includegraphics[
        width=\textwidth,
        keepaspectratio
    ]{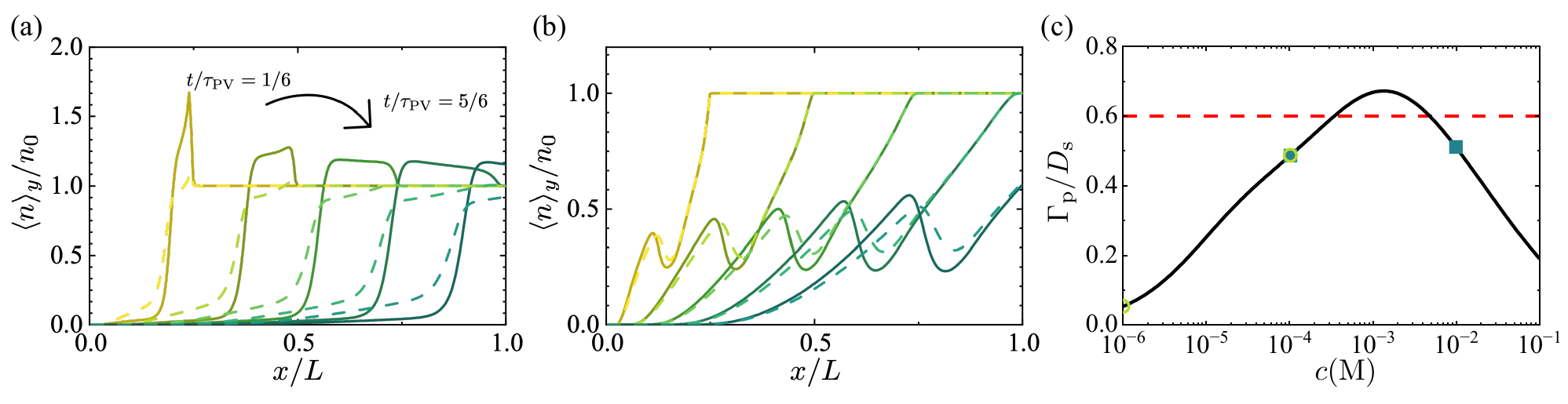}
    \captionsetup{
        width=\textwidth,
        justification=justified,
        singlelinecheck=false
    }
    \caption{
    Effect of concentration-dependent diffusiophoretic mobility on particle
    transport in a two-dimensional channel.
    (a,b) Normalized cross-sectionally averaged particle density
    \(\langle n\rangle_y/n_0\) at
    \(t/\tau_{\mathrm{PV}}
    =\{1/6,\,1/3,\,1/2,\,2/3,\,5/6\}\) for
    (a) attractive and (b) repulsive displacement. Colours progress from
    yellow to teal with increasing time. Solid curves denote the
    higher-concentration pair,
    \(0.1~\mathrm{mM}\) and \(10~\mathrm{mM}\), and dashed curves denote the
    lower-concentration pair,
    \(0.001~\mathrm{mM}\) and \(0.1~\mathrm{mM}\). In panel (a), the smaller
    concentration in each pair is \(c_{\mathrm i}\) and the larger
    concentration is \(c_{\mathrm f}\); in panel (b), these assignments are
    reversed. The concentration ratios are therefore
    \(c_{\mathrm f}/c_{\mathrm i}=100\) in panel (a) and
    \(c_{\mathrm f}/c_{\mathrm i}=10^{-2}\) in panel (b).
    (c) Dimensionless diffusiophoretic mobility
    \(\bar{\Gamma}_{\mathrm p}(c)=\Gamma_{\mathrm p}(c)/D_{\mathrm s}\)
    calculated from \eqref{eq:Gp(c)} for the representative latex--NaCl
    parameterization. Open light-green circles mark the lower-concentration
    pair and filled dark-green squares mark the higher-concentration pair.
    The symbols overlap at \(c=10^{-4}~\mathrm{M}\), which is common to both
    pairs. The red dashed line denotes the constant value
    \(\bar{\Gamma}_{\mathrm p}=0.6\) used in the main analysis.
    }
    \label{fig:CSC}
\end{figure*}

Figure~\ref{fig:CSC}(c) shows the resulting dimensionless mobility
\(\bar{\Gamma}_{\mathrm p}(c)\) and marks the two concentration pairs used in
the simulations. The predicted mobility is non-monotonic over the plotted
range, reaching a maximum near \(c=10^{-3}~\mathrm{M}\). The
higher-concentration pair,
\(10^{-4}\)--\(10^{-2}~\mathrm{M}\), samples
\(\bar{\Gamma}_{\mathrm p}\) values of approximately \(0.5\)--\(0.7\), close
to the constant value \(\bar{\Gamma}_{\mathrm p}=0.6\) used in the main
analysis. By contrast, the lower-concentration pair extends to
\(c=10^{-6}~\mathrm{M}\), where
\(\bar{\Gamma}_{\mathrm p}\simeq0.05\). It therefore probes a regime in which
the mobility is nearly an order of magnitude smaller than that used in the
constant-mobility simulations.

The corresponding particle-density profiles are shown in
figure~\ref{fig:CSC}(a,b). The largest quantitative difference occurs in the
attractive case. For the lower-concentration pair, the leading-edge density
peak is reduced and the transition is broader than for the
higher-concentration pair. This weakening of the front compression is
consistent with the substantially smaller mobility sampled on the dilute side
of the solute front. Nevertheless, both attractive cases retain a relatively
steep, compact leading front. In the repulsive case, the differences between
the two concentration ranges are less pronounced, and both cases retain the
broad trailing distribution associated with migration toward the slower
near-wall streamlines. Thus, concentration-dependent mobility modifies the
degree of particle-front sharpening and broadening but not their qualitative
character. The cross-streamline migration mechanism identified in the
constant-mobility simulations remains operative.

Within the simulated concentration range,
\(10^{-6}~\mathrm{M}\leq c\leq10^{-2}~\mathrm{M}\), allowing the
diffusiophoretic mobility to vary with the local salt concentration changes
the quantitative front shape, particularly for the dilute attractive case,
but does not alter the principal transport signatures: attractive fronts
transfer particles toward faster centreline streamlines and sharpen the
leading edge, whereas repulsive fronts transfer particles toward slower
near-wall streamlines and broaden the trailing distribution. A broader
parameter study would be required to assess regimes in which
\(\Gamma_{\mathrm p}(c)\) changes sign or
\(\lvert\Gamma_{\mathrm p}(c)\rvert/D_{\mathrm p}=O(1)\), so that particle
diffusion competes directly with diffusiophoretic migration
\citep{keh2000diffusiophoretic,prieve_diffusiophoresis_2019,
gupta_diffusiophoresis_2020,ault_physicochemical_2024}.

\section{Large-deviation formulation for two-dimensional Poiseuille flow}
\label{sec:ld-plane-poiseuille}

This appendix develops the large-deviation solute formulation used in
\S\ref{sec:ld-particle-transport}. Following
\citet{haynes2014dispersion}, we construct and validate a finite-\(\zeta\)
approximation to the solute Green's function for plane Poiseuille flow,
derive the corresponding step solution by superposition, and recover the
Taylor-core field of \S\ref{sec:Model} in the small-\(\zeta\) limit.

\subsection{Finite-\(\zeta\) large-deviation reconstruction}
\label{subsec:ld-finite-q-reconstruction}

We construct the finite-\(\zeta\) impulse response in a frame moving with the
mean flow, using the conjugate variable \(q\) to solve the associated
transverse eigenproblem. The solute concentration satisfies
\begin{equation}
  \partial_t c+u_x(y)\partial_x c
  =
  D_{\mathrm s}
  \left(
    \partial_x^2 c+\partial_y^2 c
  \right),
  \qquad
  u_x(y)
  =
  \frac{3}{2}U_{\mathrm m}
  \left(1-\frac{y^2}{a^2}\right),
  \label{eq:ld_solute_dimensional}
\end{equation}
with no-flux conditions at \(y=0\) and \(y=a\). We introduce
\begin{equation}
  \theta
  =
  \frac{c-c_{\mathrm i}}{c_{\mathrm f}-c_{\mathrm i}},
  \qquad
  \bar y=\frac{y}{a},
  \qquad
  \tilde x=\frac{x-U_{\mathrm m}t}{a},
  \qquad
  \tilde t=\frac{D_{\mathrm s}t}{a^2},
  \qquad
  \mathrm{Pe}_{\mathrm s}
  =
  \frac{U_{\mathrm m}a}{D_{\mathrm s}}.
  \label{eq:ld_nondim_variables}
\end{equation}
In the moving frame,
\begin{equation}
  \partial_{\tilde t}\theta
  +
  \mathrm{Pe}_{\mathrm s}v(\bar y)
  \partial_{\tilde x}\theta
  =
  \partial_{\tilde x}^2\theta
  +
  \partial_{\bar y}^2\theta,
  \qquad
  \partial_{\bar y}\theta=0
  \quad\text{at}\quad
  \bar y=0,1,
  \label{eq:ld_solute_dimensionless}
\end{equation}
where
\begin{equation}
  v(\bar y)
  =
  \frac{u_x-U_{\mathrm m}}{U_{\mathrm m}}
  =
  \frac12(1-3\bar y^2),
  \qquad
  \int_0^1v(\bar y)\,\mathrm d\bar y=0.
  \label{eq:ld_v_def}
\end{equation}

Consider an impulse that is initially uniform across the half-channel,
\[
  G(\tilde x,\bar y,0)=\delta(\tilde x).
\]
For \(\tilde t\gg1\), its leading large-deviation form is
\begin{equation}
  G(\tilde x,\bar y,\tilde t)
  \simeq
  \tilde t^{-1/2}
  B(\zeta)\phi_\zeta(\bar y)
  \exp[-\tilde t\,g(\zeta)],
  \qquad
  \zeta=\frac{\tilde x}{\tilde t}.
  \label{eq:ld_impulse_response}
\end{equation}
Substitution into \eqref{eq:ld_solute_dimensionless} and retention of the
leading terms as \(\tilde t\to\infty\) give
\begin{equation}
  \frac{\mathrm d^2\phi_\zeta}{\mathrm d\bar y^2}
  +
  \left[
    \mathrm{Pe}_{\mathrm s}g'(\zeta)v(\bar y)
    +
    g'(\zeta)^2
  \right]
  \phi_\zeta
  =
  \left[
    \zeta g'(\zeta)-g(\zeta)
  \right]
  \phi_\zeta,
  \qquad
  \frac{\mathrm d\phi_\zeta}{\mathrm d\bar y}=0
  \quad\text{at}\quad
  \bar y=0,1.
  \label{eq:ld_wkb_leading}
\end{equation}

To compute this solution, define the variable \(q\) conjugate to the ray
velocity \(\zeta\), together with the Legendre transform of \(g\):
\begin{equation}
  q
  =
  g'(\zeta;\mathrm{Pe}_{\mathrm s}),
  \qquad
  f(q;\mathrm{Pe}_{\mathrm s})
  =
  q\zeta-g(\zeta;\mathrm{Pe}_{\mathrm s}).
  \label{eq:ld_q_f_definition}
\end{equation}
The inverse relations are
\[
  \zeta=f'(q),
  \qquad
  g(\zeta)=qf'(q)-f(q).
\]
Defining
\begin{equation}
  \mathcal L_q
  =
  \frac{\mathrm d^2}{\mathrm d\bar y^2}
  +
  \mathrm{Pe}_{\mathrm s}qv(\bar y)
  +
  q^2,
  \label{eq:ld_Lq_def}
\end{equation}
equation \eqref{eq:ld_wkb_leading} becomes
\begin{equation}
  \mathcal L_q\phi_q
  =
  f(q;\mathrm{Pe}_{\mathrm s})\phi_q,
  \qquad
  \frac{\mathrm d\phi_q}{\mathrm d\bar y}=0
  \quad\text{at}\quad
  \bar y=0,1.
  \label{eq:ld_finiteq_evp}
\end{equation}
We use the principal eigenvalue and choose its positive eigenfunction with
normalization
\begin{equation}
  \int_0^1\phi_q^2(\bar y)\,\mathrm d\bar y=1.
  \label{eq:ld_phi_normalization}
\end{equation}

On this principal branch, \(\zeta=f'(q)\) is one-to-one
\citep{haynes2014dispersion}. Thus \(q\) is only a convenient
parametrization of the finite-\(\zeta\) solution, with
\(\phi_q=\phi_{\zeta(q)}\). For an impulse uniform across the channel, the
normal-mode projection and steepest-descent approximation give
\citep{haynes2014dispersion}
\begin{equation}
  B(q)
  \equiv
  B(\zeta(q))
  =
  \frac{\mathcal A(q)}
       {\sqrt{2\pi f''(q)}},
  \qquad
  \mathcal A(q)
  =
  \int_0^1\phi_q(\bar y_0)\,\mathrm d\bar y_0.
  \label{eq:ld_prefactor_B}
\end{equation}
Hence the impulse response can be evaluated directly from
\[
  G(\tilde x,\bar y,\tilde t)
  \simeq
  \frac{\mathcal A(q)\phi_q(\bar y)}
       {\sqrt{2\pi\tilde t\,f''(q)}}
  \exp\left\{
    -\tilde t\left[qf'(q)-f(q)\right]
  \right\},
  \qquad
  f'(q)=\frac{\tilde x}{\tilde t}.
\]
The dependence of \(f\), \(g\), \(\phi_q\), and \(B\) on
\(\mathrm{Pe}_{\mathrm s}\) is suppressed below.

We validate this reconstruction against a direct numerical solution of
\eqref{eq:ld_solute_dimensionless}. The direct solution uses a Fourier
discretization in \(\tilde x\) and a converged high-order discretization in
\(\bar y\) satisfying the homogeneous Neumann conditions. We take
\(\mathrm{Pe}_{\mathrm s}=2000\), for which the impulse-response tails are
strongly non-Gaussian.

\begin{figure*}
  \centering
  \includegraphics[width=\textwidth,keepaspectratio]{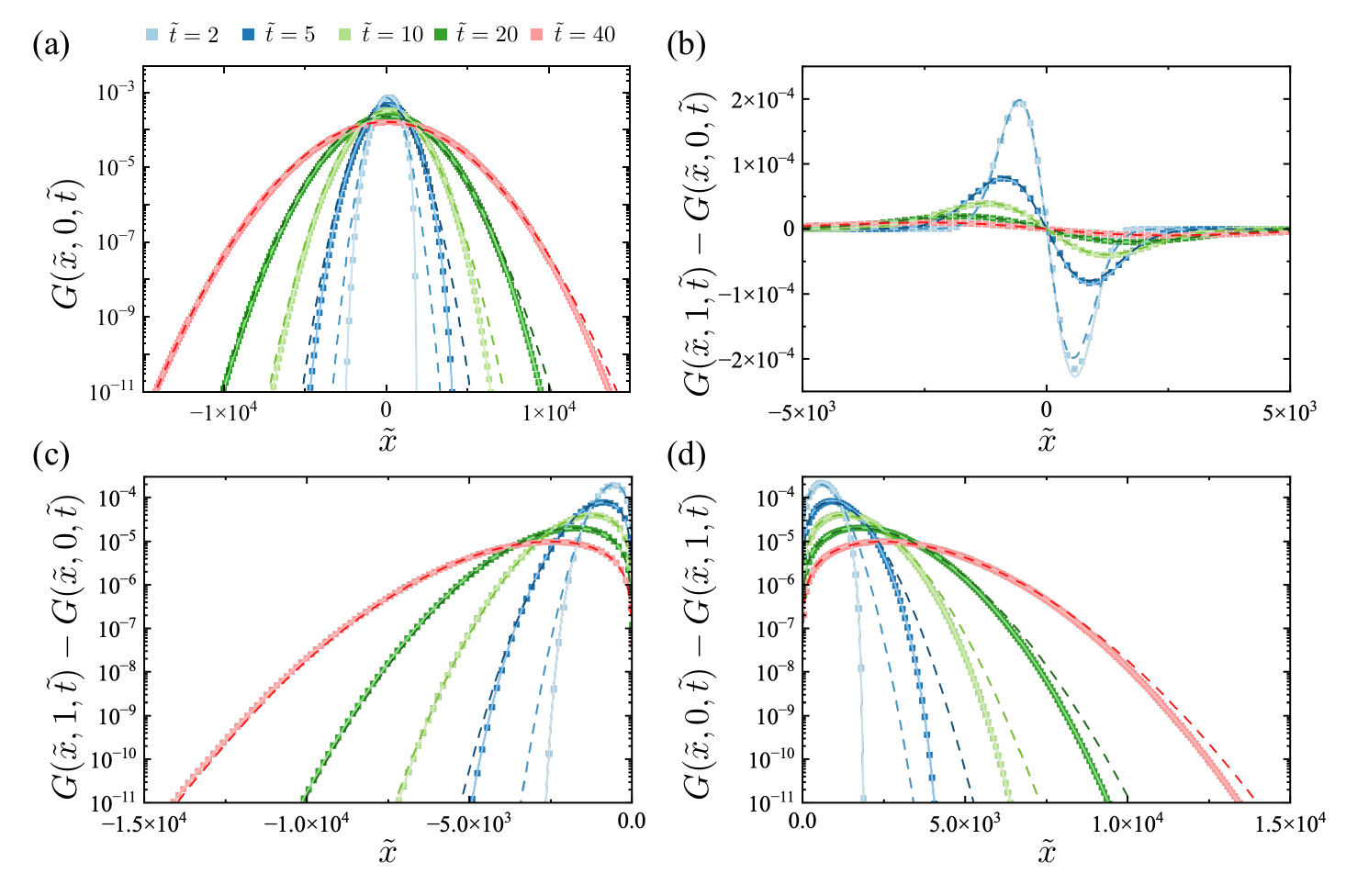}
  \captionsetup{
    width=\textwidth,
    justification=justified,
    singlelinecheck=false
  }
  \caption{
  Validation of the large-deviation Green's function for a
  cross-sectionally uniform impulse in plane Poiseuille flow at
  \(\mathrm{Pe}_{\mathrm s}=2000\). Symbols denote direct numerical
  solutions, solid lines denote the finite-\(q\) reconstruction
  \eqref{eq:ld_impulse_response}, and dashed lines denote the Taylor-core
  approximation. Colours correspond to
  \(\tilde t=\{2,5,10,20,40\}\).
  (a) Centreline concentration \(G(\tilde x,0,\tilde t)\).
  (b) Wall--centreline difference
  \(G(\tilde x,1,\tilde t)-G(\tilde x,0,\tilde t)\).
  (c) The same difference in the negative-\(\tilde x\) tail.
  (d) Centreline--wall difference
  \(G(\tilde x,0,\tilde t)-G(\tilde x,1,\tilde t)\) in the
  positive-\(\tilde x\) tail.
  Logarithmic scales are used in panels (a), (c), and (d). In panel (a),
  the Taylor approximation is the leading Gaussian \(G_0\) in
  \eqref{eq:theta0_green_ld}; in panels (b--d), it includes the first
  transverse correction in
  \eqref{eq:first_order_green_function_explicit}.
  }
  \label{fig:Green_Validation}
\end{figure*}

Figure~\ref{fig:Green_Validation} shows that the finite-\(q\)
reconstruction captures both the streamwise concentration and its transverse
variation. The Taylor approximation is accurate in the distribution core,
where \(\tilde x=O(\tilde t^{1/2})\), but departs from the non-Gaussian
tails because it retains only the quadratic, small-\(q\) approximation to
\(f(q)\).

For the displacement problem,
\begin{equation}
  \theta(\tilde x,\bar y,0)=H(-\tilde x).
\end{equation}
By linearity, the step solution is the cumulative impulse response:
\begin{equation}
  \theta_{\mathrm{LD}}(\tilde x,\bar y,\tilde t)
  =
  \int_{-\infty}^{0}
  G(\tilde x-\tilde x_0,\bar y,\tilde t)\,
  \mathrm d\tilde x_0
  =
  \int_{\tilde x}^{\infty}
  G(r,\bar y,\tilde t)\,\mathrm dr.
  \label{eq:ld_step_convolution}
\end{equation}
Along the integration variable \(r\),
\[
  \frac{r}{\tilde t}=f'(q),
  \qquad
  r=\tilde t f'(q),
  \qquad
  \mathrm dr=\tilde t f''(q)\,\mathrm dq.
\]
Let \(q_{\tilde x}\) satisfy
\begin{equation}
  f'(q_{\tilde x})
  =
  \frac{\tilde x}{\tilde t}.
  \label{eq:q_x_definition}
\end{equation}
Substitution of \eqref{eq:ld_impulse_response} and
\eqref{eq:ld_prefactor_B} into \eqref{eq:ld_step_convolution} then gives
\begin{equation}
  \theta_{\mathrm{LD}}(\tilde x,\bar y,\tilde t)
  \simeq
  \int_{q_{\tilde x}}^{\infty}
  \left[
    \frac{\tilde t f''(q)}{2\pi}
  \right]^{1/2}
  \mathcal A(q)\phi_q(\bar y)
  \exp\left\{
    -\tilde t
    \left[
      qf'(q)-f(q)
    \right]
  \right\}
  \,\mathrm dq.
  \label{eq:ld_step_q_integral}
\end{equation}
Numerically, the upper limit is replaced by the largest tabulated value of
\(q\), chosen so that the omitted contribution is exponentially small.

Finally, the dimensional concentration is
\begin{equation}
  c_{\mathrm{LD}}(x,y,t)
  =
  c_{\mathrm i}
  +
  (c_{\mathrm f}-c_{\mathrm i})
  \theta_{\mathrm{LD}}(\tilde x,\bar y,\tilde t),
  \qquad
  \tilde x=\frac{x-U_{\mathrm m}t}{a},
  \quad
  \bar y=\frac{y}{a},
  \quad
  \tilde t=\frac{D_{\mathrm s}t}{a^2}.
  \label{eq:c_LD_dimensional}
\end{equation}
Its transverse gradient is
\begin{equation}
  \frac{\partial c_{\mathrm{LD}}}{\partial y}
  =
  \frac{c_{\mathrm f}-c_{\mathrm i}}{a}
  \frac{\partial\theta_{\mathrm{LD}}}{\partial\bar y}.
  \label{eq:cy_LD_dimensional}
\end{equation}
In \eqref{eq:ld_step_q_integral},
\(\partial_{\bar y}\theta_{\mathrm{LD}}\) is obtained simply by replacing
\(\phi_q(\bar y)\) with \(\phi_q'(\bar y)\).

\subsection{Small-\(\zeta\) limit and Taylor-core solute field}
\label{subsec:ld-taylor-core}

The Taylor regime corresponds to the core of the distribution,
\begin{equation}
  \tilde x=O(\tilde t^{1/2}),
  \qquad
  \zeta=\frac{\tilde x}{\tilde t}
  =O(\tilde t^{-1/2}),
  \qquad
  q=O(\tilde t^{-1/2}).
  \label{eq:taylor_core_scaling}
\end{equation}
We therefore expand the principal eigenpair about \(q=0\):
\begin{align}
  \phi_q(\bar y)
  &=
  \phi_0(\bar y)
  +
  q\phi_1(\bar y)
  +
  q^2\phi_2(\bar y)
  +
  O(q^3),
  \label{eq:phi_smallq_expansion}
  \\
  f(q)
  &=
  \alpha_0+\alpha_1q+\alpha_2q^2+O(q^3).
  \label{eq:f_alpha_smallq_expansion}
\end{align}

Substitution into \eqref{eq:ld_finiteq_evp} gives, at \(O(1)\), the
normalized principal eigenpair
\begin{equation}
  \alpha_0=0,
  \qquad
  \phi_0(\bar y)=1.
  \label{eq:phi0_alpha0_solution_ld}
\end{equation}
At \(O(q)\),
\begin{equation}
  \phi_1''
  +
  \mathrm{Pe}_{\mathrm s}v(\bar y)
  =
  \alpha_1,
  \qquad
  \phi_1'(0)=\phi_1'(1)=0.
  \label{eq:phi1_problem_ld}
\end{equation}
Because \(v\) has zero cross-sectional mean, integration gives
\(\alpha_1=0\). Differentiating the normalization condition at \(q=0\)
further gives
\[
  \int_0^1\phi_1(\bar y)\,\mathrm d\bar y=0.
\]

Define
\begin{equation}
  I(\bar y)
  =
  \int_0^{\bar y}v(s)\,\mathrm ds
  =
  \frac12(\bar y-\bar y^3),
  \label{eq:I_def_ld}
\end{equation}
and
\begin{equation}
  \chi(\bar y)
  =
  \int_0^{\bar y}I(s)\,\mathrm ds
  =
  \frac18\bar y^2(2-\bar y^2),
  \qquad
  \bar\chi
  =
  \int_0^1\chi(\bar y)\,\mathrm d\bar y
  =
  \frac{7}{120}.
  \label{eq:chi_def_ld}
\end{equation}
The first eigenfunction correction is then
\begin{equation}
  \phi_1(\bar y)
  =
  -\mathrm{Pe}_{\mathrm s}
  \left[
    \chi(\bar y)-\bar\chi
  \right].
  \label{eq:phi1_solution_ld}
\end{equation}

At \(O(q^2)\), integration of the eigenvalue equation across the channel
gives
\begin{align}
  \alpha_2
  &=
  1
  +
  \mathrm{Pe}_{\mathrm s}
  \int_0^1v(\bar y)\phi_1(\bar y)\,\mathrm d\bar y
  \nonumber\\
  &=
  1
  +
  \mathrm{Pe}_{\mathrm s}^2
  \int_0^1I^2(\bar y)\,\mathrm d\bar y
  =
  1+\frac{2}{105}\mathrm{Pe}_{\mathrm s}^2,
  \label{eq:alpha2_poisseuille_ld}
\end{align}
where integration by parts uses \(v=I'\) and \(I(0)=I(1)=0\).

The quadratic eigenvalue coefficient is the dimensionless Taylor
diffusivity \citep{haynes2014dispersion}:
\begin{equation}
  \bar D_{\mathrm{s,eff}}
  =
  \frac12f''(0)
  =
  \alpha_2
  =
  1+\frac{2}{105}\mathrm{Pe}_{\mathrm s}^2.
  \label{eq:Dbar_alpha2_ld}
\end{equation}
Thus
\begin{equation}
  f(q)
  =
  \bar D_{\mathrm{s,eff}}q^2+O(q^3),
  \label{eq:f_smallq_poisseuille}
\end{equation}
and the corresponding dimensional diffusivity is
\begin{equation}
  D_{\mathrm{s,eff}}
  =
  D_{\mathrm s}\bar D_{\mathrm{s,eff}}
  =
  D_{\mathrm s}
  \left(
    1+\frac{2}{105}\mathrm{Pe}_{\mathrm s}^2
  \right).
  \label{eq:Ds_eff_from_ld}
\end{equation}

The Legendre relations now give
\begin{equation}
  \zeta
  =
  f'(q)
  =
  2\bar D_{\mathrm{s,eff}}q+O(q^2),
  \qquad
  q(\zeta)
  =
  \frac{\zeta}{2\bar D_{\mathrm{s,eff}}}
  +O(\zeta^2),
  \label{eq:zeta_smallq_ld}
\end{equation}
and
\begin{equation}
  g(\zeta)
  =
  \frac{\zeta^2}{4\bar D_{\mathrm{s,eff}}}
  +O(\zeta^3).
  \label{eq:g_smallzeta_poisseuille}
\end{equation}
Consequently, in the Taylor core,
\begin{equation}
  \tilde t\,
  g\!\left(\frac{\tilde x}{\tilde t}\right)
  =
  \frac{\tilde x^2}
       {4\bar D_{\mathrm{s,eff}}\tilde t}
  +
  O(\tilde t^{-1/2}).
  \label{eq:ld_core_exponent}
\end{equation}

Since
\[
  \phi_0=1,
  \qquad
  \mathcal A(0)=1,
  \qquad
  f''(0)=2\bar D_{\mathrm{s,eff}},
\]
the prefactor at \(q=0\) is
\begin{equation}
  B(0)
  =
  \frac{1}{\sqrt{4\pi\bar D_{\mathrm{s,eff}}}}.
  \label{eq:B0_ld}
\end{equation}
The leading Taylor-core Green's function is therefore
\begin{equation}
  G_0(\tilde x,\tilde t)
  =
  \frac{1}
       {\sqrt{4\pi\bar D_{\mathrm{s,eff}}\tilde t}}
  \exp\left[
    -\frac{\tilde x^2}
          {4\bar D_{\mathrm{s,eff}}\tilde t}
  \right].
  \label{eq:theta0_green_ld}
\end{equation}

The first transverse correction follows from
\begin{equation}
  \phi_q(\bar y)
  =
  1
  -
  q\,\mathrm{Pe}_{\mathrm s}
  \left[
    \chi(\bar y)-\bar\chi
  \right]
  +
  O(q^2).
  \label{eq:phi_q_smallq_explicit}
\end{equation}
In the core,
\[
  q
  =
  \frac{\tilde x}
       {2\bar D_{\mathrm{s,eff}}\tilde t}
  +
  O(\tilde t^{-1}),
\]
and hence
\begin{equation}
  qG_0
  =
  -\frac{\partial G_0}{\partial\tilde x}
  +
  O(\tilde t^{-1})G_0.
  \label{eq:qG0_identity_ld}
\end{equation}
The remaining first-order corrections from the prefactor and the
\(O(q^3)\) term in \(f\) are independent of \(\bar y\); denote their sum by
\(G_{1a}\). The term proportional to \(\bar\chi\) is also independent of
\(\bar y\) and may be absorbed into \(G_{1a}\). Thus
\begin{equation}
  G(\tilde x,\bar y,\tilde t)
  =
  G_0(\tilde x,\tilde t)
  +
  \mathrm{Pe}_{\mathrm s}\chi(\bar y)
  \frac{\partial G_0}{\partial\tilde x}
  +
  G_{1a}(\tilde x,\tilde t)
  +
  O(\tilde t^{-1})G_0.
  \label{eq:first_order_green_function_explicit}
\end{equation}

The step solution is the cumulative Green's function. Its leading,
cross-sectionally uniform part is
\begin{align}
  X(\tilde x,\tilde t)
  &=
  \int_{\tilde x}^{\infty}
  G_0(r,\tilde t)\,\mathrm dr
  \nonumber\\
  &=
  \frac12
  \erfc\left(
    \frac{\tilde x}
         {\sqrt{4\bar D_{\mathrm{s,eff}}\tilde t}}
  \right)
  \nonumber\\
  &=
  \frac12
  \left[
    1-
    \erf\left(
      \frac{x-U_{\mathrm m}t}
           {\sqrt{4D_{\mathrm{s,eff}}t}}
    \right)
  \right].
  \label{eq:Xfunction_from_ld}
\end{align}
Moreover,
\begin{equation}
  \int_{\tilde x}^{\infty}
  \frac{\partial G_0}{\partial r}(r,\tilde t)\,
  \mathrm dr
  =
  -G_0(\tilde x,\tilde t)
  =
  \frac{\partial X}{\partial\tilde x}
  =
  a\frac{\partial X}{\partial x}.
  \label{eq:step_derivative_identity_ld}
\end{equation}
The cross-sectionally varying Taylor field is therefore
\begin{equation}
  \theta_{\mathrm T}(\tilde x,\bar y,\tilde t)
  =
  X(\tilde x,\tilde t)
  +
  \mathrm{Pe}_{\mathrm s}\chi(\bar y)
  \frac{\partial X}{\partial\tilde x},
  \qquad
  \chi(\bar y)
  =
  \frac18\bar y^2(2-\bar y^2).
  \label{eq:ld_theta_T_particle}
\end{equation}

Let
\[
  \Theta_{1a}(x,t)
  =
  \int_{\tilde x}^{\infty}
  G_{1a}(r,\tilde t)\,\mathrm dr,
\]
which is independent of \(y\). The dimensional concentration is then
\begin{equation}
  c(x,y,t)
  =
  c_{\mathrm i}
  +
  (c_{\mathrm f}-c_{\mathrm i})
  \left[
    \theta_{\mathrm T}
    +
    \Theta_{1a}(x,t)
    +
    O(\tilde t^{-1})
  \right].
  \label{eq:soluteFieldExpr_from_ld_compact}
\end{equation}
Equivalently,
\begin{equation}
  c(x,y,t)
  =
  c_{\mathrm i}
  +
  (c_{\mathrm f}-c_{\mathrm i})
  \left\{
    X(x,t)
    +
    \frac{\mathrm{Pe}_{\mathrm s}}{8}
    \left(\frac{y}{a}\right)^2
    \left[
      2-\left(\frac{y}{a}\right)^2
    \right]
    a\frac{\partial X}{\partial x}
    +
    \Theta_{1a}(x,t)
    +
    O(\tilde t^{-1})
  \right\}.
  \label{eq:soluteFieldExpr_from_ld}
\end{equation}

Because \(\Theta_{1a}\) is independent of \(y\), it does not contribute to
the transverse gradient. Differentiation therefore gives
\begin{equation}
  \frac{\partial c}{\partial y}
  =
  (c_{\mathrm f}-c_{\mathrm i})
  \frac{\mathrm{Pe}_{\mathrm s}}{2}
  \frac{y}{a}
  \left[
    1-\left(\frac{y}{a}\right)^2
  \right]
  \frac{\partial X}{\partial x}
  +
  \frac{c_{\mathrm f}-c_{\mathrm i}}{a}
  O(\tilde t^{-1}).
  \label{eq:solute_transverse_gradient_from_ld}
\end{equation}
This recovers \eqref{eq:solute_transverse_gradient} and shows that the
finite-\(q\) reconstruction reduces, in the distribution core, to the
Taylor-regime solute field used in \S\ref{sec:Model}.

\section{Numerical implementation of the channel simulations}
\label{app:numerical_methods}

The Eulerian and Lagrangian simulations described in
\S\ref{sec:Simulation} were performed using custom solvers within the
OpenFOAM finite-volume framework. The two-dimensional half-channel domain,
\(0\leq x\leq L\) and \(0\leq y\leq a\), had
\(L=30\,\mathrm{mm}\) and \(a=20\,\mu\mathrm{m}\). It was discretized by a
uniform orthogonal mesh with
\(\Delta x=\Delta y=1\,\mu\mathrm{m}\), giving
\(30000\times20=6\times10^5\) control volumes. The inactive third direction
contained one cell with empty boundary conditions. A steady pressure-driven
flow with cross-sectional mean
\(U_{\mathrm m}=100\,\mu\mathrm{m\,s^{-1}}\) was first computed using
\texttt{simpleFoam} and then held fixed. The resulting velocity profile agreed
with the plane-Poiseuille solution in \eqref{eq:poiseuille_sim}.

In the Eulerian solver, the solute and particle equations
\eqref{eq:c_ade} and \eqref{eq:n_ade} were advanced sequentially. The
concentration and particle density were normalized by \(c_{\mathrm i}\) and
\(n_0\), respectively, so that the initial and inlet solute concentrations
were \(1\) and \(\beta=c_{\mathrm f}/c_{\mathrm i}\). After each solute
update, the diffusiophoretic velocity was evaluated as
\begin{equation}
\mathbf{u}_{\mathrm{DP}}
=
\Gamma_{\mathrm p}
\frac{\nabla c}{\max(c,c_{\min})},
\qquad
c_{\min}=10^{-12},
\label{eq:numerical_udp}
\end{equation}
where the floor, well below the minimum simulated concentration, prevents
division by zero. The corresponding face flux was included conservatively in
the particle-density equation.

Time integration used an off-centred Crank--Nicolson scheme with coefficient
\(0.9\). Convective fluxes, including diffusiophoretic particle transport,
were discretized using the bounded van Leer scheme; diffusive fluxes used
linear central interpolation, and \(\nabla c\) was evaluated using a limited
linear Gauss scheme. The time step was
\(\Delta t=10^{-3}\,\mathrm{s}\), giving a maximum flow Courant number of
approximately \(0.15\). The scalar systems were solved using a preconditioned
bi-conjugate-gradient method with absolute tolerances of \(10^{-8}\) for
\(c\) and \(10^{-10}\) for \(n\). At the inlet, \(c=\beta\) and \(n=0\);
zero streamwise gradients were imposed at the advective outlet. Symmetry was
applied at \(y=0\), and zero normal solute and particle fluxes at \(y=a\).

The Lagrangian calculations used the same mesh, fixed flow field, solute
solver, and boundary conditions. Before particle release, the solute was
evolved for
\(t_{\mathrm{pre}}=20\,\mathrm{s}=50\tau_{\mathrm s}\) with
\(\Delta t=10^{-3}\,\mathrm{s}\). Each protocol then used
\(5\times10^4\) point particles, uniformly distributed across the half-gap
and over
\[
5\,\mathrm{mm}\leq x\leq30\,\mathrm{mm}.
\]
Thus, even the particles nearest the inlet were initially several
Taylor-front widths ahead of the solute transition and experienced
negligible diffusiophoretic drift.

During tracking, the solute continued to evolve and particles were advanced
according to \eqref{eq:lagrangian_trajectory} with \(D_{\mathrm p}=0\).
Flow and diffusiophoretic velocities were interpolated by cell--point
interpolation. Particle trajectories were integrated explicitly using
OpenFOAM face walking, with substeps chosen to keep the particle Courant
number below \(0.25\); the global time step was
\(2\times10^{-3}\,\mathrm{s}\). Particles were reflected at the impermeable
wall and removed at the outlet, with no inlet injection or Brownian
displacements. Times reported in figure~\ref{fig:LPT} include the
\(20\,\mathrm{s}\) solute pre-evolution period.

\bibliographystyle{jfm}
\bibliography{jfm}

\end{document}